\documentclass[a4paper,aps,secnumarabic, twocolumn,
  balancelastpage,amsmath,amssymb,nofootinbib,
  preprint,10pt
]{revtex4-2}

\usepackage{graphicx}% Include figure files
\usepackage{dcolumn}% Align table columns on decimal point
\usepackage{bm}% bold math
\usepackage{float}
\usepackage[colorlinks=true, urlcolor=blue, linkcolor=red]{hyperref}
\usepackage[dvipsnames]{xcolor}

\usepackage{chemformula}
\usepackage[margin=1.8cm]{geometry}
\usepackage{soul}

\begin{document}

\title{Elf autoencoder: unsupervised exploration of flat-band materials using electronic band structure fingerprints}

\author{Henry Kelbrick Pentz}
\thanks{These authors contributed equally}
\author{Thomas Warford}
\thanks{These authors contributed equally}
\author{Ivan Timokhin}
\author{Qian Yang}
\email[]{qian.yang@manchester.ac.uk }
\author{Anupam Bhattacharya}
\email[]{anupamcounting@gmail.com}
\author{Artem Mishchenko}
\email[]{artem.mishchenko@manchester.ac.uk}
\affiliation{Department of Physics and Astronomy, University of Manchester}

\begin{abstract}
Two-dimensional materials with flat electronic bands are promising for realizing exotic quantum phenomena such as unconventional superconductivity and nontrivial topology, but exploring their vast chemical space remains challenging. Here, we introduce an unsupervised convolutional autoencoder agent (elf) that operates on electronic band structure images and is capable of mapping band features and extracting the latent space representation as a fingerprint, enabling autonomous clustering of materials with common electronic properties beyond traditional chemical paradigms. Unsupervised visualisation of the latent space then helps to uncover hidden chemical trends and identify promising candidates based on similarities to well-studied exemplars. Our framework paves the way for the accelerated discovery of novel flat-band materials with desirable electronic characteristics. It complements high-throughput ab initio methods by rapidly screening candidates and guides further investigations into the mechanisms governing the emergence of flat-band physics. We believe the elf autoencoder will be a valuable tool for the autonomous discovery of previously unexplored flat-band materials, aiding in the unbiased identification of compounds with desirable electronic properties in vast 2D chemical space. 
\end{abstract}

\maketitle

\section{Introduction} \label{I}
High-throughput computational methods based on machine learning are quickly becoming the paradigm approach to next-generation materials discovery \cite{DEEP_1,rashid2024review, leeman2024challenges, jain2024machine, lyngby2024ab, cheetham2024artificial, alverson2024generative, adam2024automated, wang2023scientific, szymanski2023autonomous}. In this context, testing the stability and properties of the vast potential materials at the lab scale poses a significant bottleneck. Thus, AI-based approaches to classify patterns among identified materials via their characteristic fingerprints (machine-learnable vector forms of material properties) are urgently needed. These techniques are crucial for understanding the emergent properties in computationally generated materials, such as flat band formation \cite{Anupam}, band topology, superconductivity, photovoltaic potential, catalytic behaviour, and more. They also allow predicted materials to be linked to compounds with experimentally confirmed properties. Using this approach, large sets of predicted materials can now be analyzed at once, allowing candidates with the most promising properties to be easily flagged for further investigation.

This paper presents a novel approach to generating materials fingerprints based solely on their electronic band structure, sourced from precomputed databases generated via Density Functional Theory (DFT). Our goal is to develop an unsupervised machine learning framework that can accelerate the discovery of flat-band materials with desirable electronic properties by enabling unbiased exploration of vast chemical space without reliance on labelled data. 
 Fingerprints based on band structures have been attempted before for exploring the vast materials space \cite{feature_2, feature_1}, but never directly with unsupervised, machine-learned features of the electronic bands. Here, we apply a convolutional autoencoder (CAE) - elf, to generate such fingerprints autonomously, facilitating the clustering of materials by electronic band features, without any bias from crystal structure. Our approach emerges as a three-fold alternative to traditional structure-based fingerprints like CrystalNNFingerprint \cite{Crystal}, allowing (1)  detecting duplicates, (2) elucidating chemical patterns among the input materials, and (3) clustering by materials' emergent electronic properties. 

Our high-throughput fingerprint generation method is easily scalable to any database containing band structure data and can be used as a complementary tool in analyzing the growing sets of computationally generated materials. The electronic fingerprints enable rapid screening of candidates identified via ab initio methods, accelerating the discovery process.  In this work, we apply this method to two-dimensional (2D) materials hosting flat electronic bands, which were determined in Ref. \cite{Anupam} using the 2Dmatpedia \cite{2Dmatpedia}, currently the largest open 2D materials database.

Flat bands, which are momentum states with approximately the same energy, have attracted considerable attention for hosting exotic strongly correlated physics \cite{checkelsky2024flat, leykam2018artificial, torma2022superconductivity, rhim2021singular}. In two dimensions, 'plane flat bands' extend in both $k_{x}$ and $k_{y}$ directions, forming an extended planar manifold in reciprocal space. The resulting quenched kinetic energy makes flat bands conducive to interactions among electrons. This has been shown to enable phenomena such as chiral plasmons \cite{plasmons}, unconventional superconductivity, originally observed in twisted bilayer graphene \cite{twist}, Chern insulator states \cite{Chern}, and much more.

We employ unsupervised clustering algorithms to analyze the fingerprints of 2D flat-band materials, from which novel chemical and electronic feature groups emerge, extending the known flat-band paradigm. Two-dimensional embedded clustering plots are used to display the chemical and band-feature groups, providing a map for the chemistry of 2D flat-band materials. Our hybrid approach also serves as a blueprint for high-throughput analysis of electronic and chemical properties of material databases. It is generalizable to any subsets of compounds with certain properties. The robustness of elf CAE band feature encoding allows strong predictions to be made for the emergent electronic properties of grouped materials, particularly when they are accompanied by well-studied compounds in the same cluster. This is achieved at an extremely low computational cost, facilitating simple adoption and generalization of our approach.

\begin{figure*}
\includegraphics[width=0.8\textwidth]{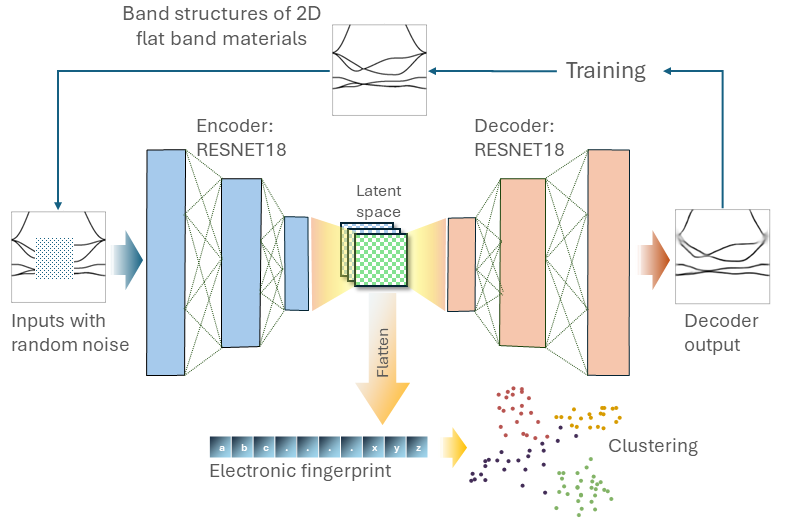}
\caption{\label{fig:pipeline} Schematic of the elf network and fingerprint analysis pipeline for autoencoded band structure fingerprints. The convolutional autoencoder is trained to reproduce electronic band structure images within an energy range of $\pm$4 eV relative to the Fermi level. The process involves encoding these images into a compressed latent space representation. This compressed representation serves as the material fingerprint. The diagram outlines the steps of training the network, applying random noise to input images during training to enhance the learning of robust band structure features, and using the encoded representation for clustering materials based on their electronic properties. The reconstructed band structure of 2dm-1 (\ch{IrF2}) is shown as an example, demonstrating the network's ability to accurately capture and reproduce band structures even in the presence of noise.}
\end{figure*}

\section{Results} \label{II}
Several attempts were recently reported to detect flat band materials in databases using computational screening and data mining techniques, creating repositories of well-documented 2D and 3D candidates \cite{Anupam, 2dnewcatelogue, crystalnet,flat_Cat, zhang2023physically}. Building on these efforts, we started this work from one such pool of 2127 flat band materials identified by Bhattacharya et al.\cite{Anupam} using the 2Dmatpedia database\cite{2Dmatpedia}. For each of the flat-band materials, the band structure image data were encoded by a trained CAE elf, and the latent space representation was flattened to produce a 98-dimensional fingerprint vector. The training and subsequent fingerprint extraction process is shown in Figure~\ref{fig:pipeline}, with random Gaussian noise applied only during the training. We employed a ResNet18 architecture for both the encoder and decoder of elf, as its deep structure allows learning nuanced features of band structure images in comparison to shallow networks. Several ResNet models with different input image sizes and latent space dimensions were tested. Optimum performance and accuracy were obtained for the chosen model with $(224\times224)$ input image size and $(7\times7\times 2$channels) latent space dimensions (More details in Section \ref{M.1}). Furthermore, by introducing random noise to the regions of input images during training, the network is forced to learn physically sensible connections between electronic state lines in the `noised' regions, relying only on the shapes of the surrounding bands. This more challenging task helps to prevent overfitting by encouraging the learning of more general band structure features that are robust to small perturbations. 

The set of fingerprints generated by the trained elf was then clustered using Hierarchical Density-Based Spatial Clustering of Applications with Noise (HDBSCAN) \cite{HDB}. The optimal parameters for HDBSCAN were determined using the optimization procedure outlined in the Methods section \ref{M.2}. Specifically, the minimum cluster size ($N_{c}$) and the minimum sample size ($N_{s}$) were set to 5 and 2, respectively. This clustering process identified 50 distinct clusters, while 1662 materials remained unclassified. 

\begin{figure*}
\includegraphics[width=\textwidth]{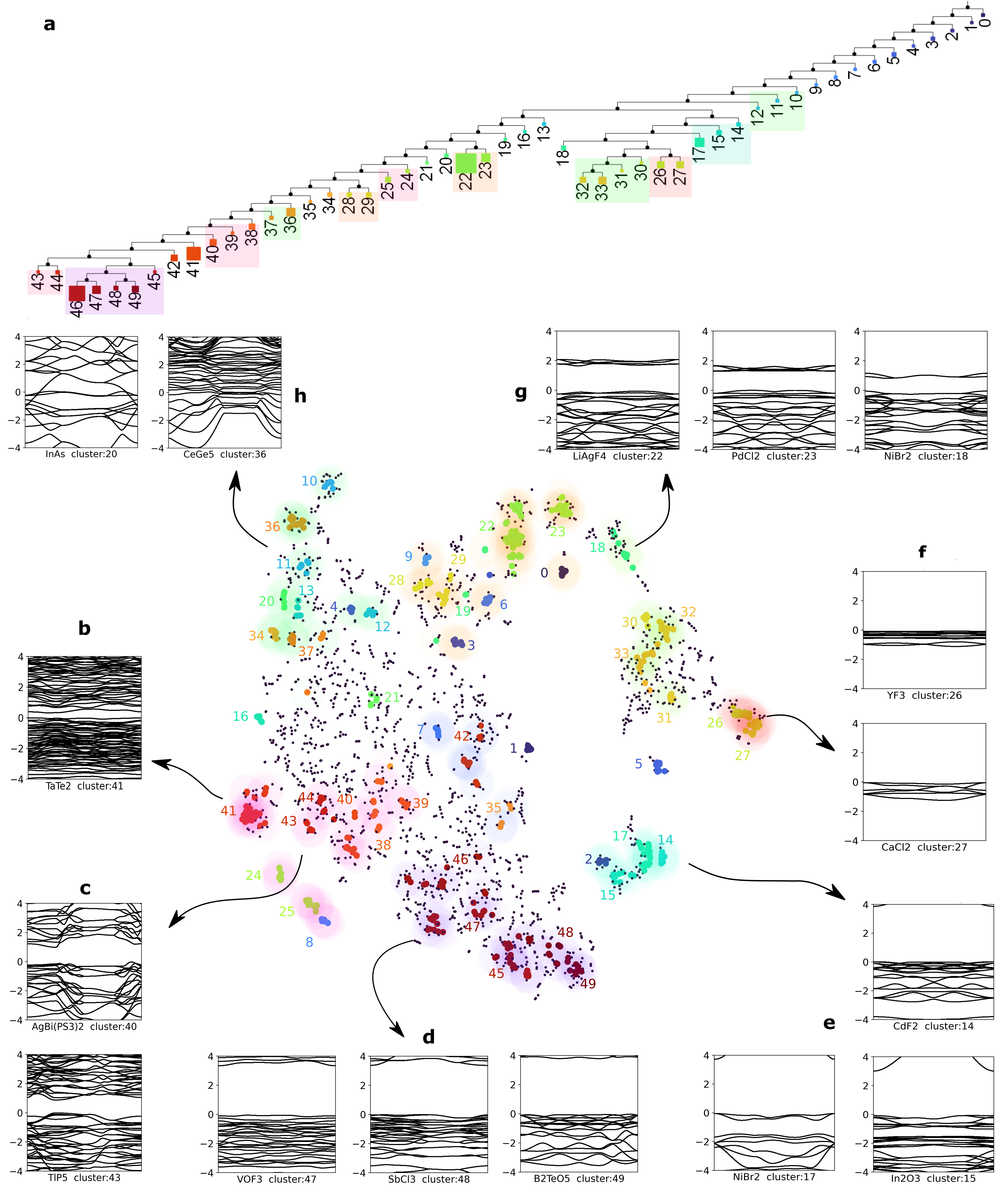}
\caption{\label{fig:tSNE} Visualization of the electronic fingerprint space and emergent chemical trends using HDBSCAN and DBSCAN clustering algorithms combined with t-SNE dimensionality reduction. \textbf{a}, Phylogenetic tree of the HDBSCAN clusters showing the relative sizes of the clusters. Clusters with similar defining features are grouped within the same coloured boxes. \textbf{b-h}, examples of band structures mapped onto t-SNE plot for the machine-learned fingerprints, with perplexity = 10, and early exaggeration = 12. The clusters determined by HDBSCAN are marked with an opaque colour, while the transparent shades from DBSCAN identify global trends and major types of band structures. The band structures of a few exemplars from a group of clusters are highlighted nearby to visualise their distinctive features: \textbf{b}, Post-transition metal compounds and transition metal chalcogenides, with vanishing, indirect, or band overlap materials (potential semi-metallic phases). \textbf{c}, 1-2 eV band gap materials with frequent crossings. \textbf{d}, Wide band gap ($>$3 eV) semiconductors. \textbf{e}, Insulators and large indirect band gap materials. \textbf{f}, Metal oxides and halides. Insulators with strips of plane flat bands at the Fermi level. \textbf{g}, Metal and transition metal halides with dispersive valence bands and plane-flat conduction bands. Within these groups, the band gap decreases from left (cluster 22) to right (cluster 18).  \textbf{h}, flat bands with metallic band structures.}
\end{figure*}

To visualize the distribution of materials in the machine-learned fingerprint space, we employed t-distributed Stochastic Neighbor Embedding (t-SNE) \cite{t-sne}. The t-SNE algorithm was applied with a perplexity value of 10 and an early exaggeration parameter of 12, to reduce the high-dimensional fingerprint space to a 2D representation. Figure~\ref{fig:tSNE} presents the t-SNE chart, displaying the clustered materials in the reduced-dimensional space. 

To further identify major groups of materials based on their electronic structure similarities, we applied an additional layer of clustering using the density-based (DBSCAN) algorithm\cite{DBSCAN}. This second clustering step was performed on the t-SNE projected coordinates of the materials, excluding those that were left unclassified by HDBSCAN. The DBSCAN parameters were set to a maximum nearest neighbour distance ($\epsilon$) on the t-SNE projected plane of 25 and minimum cluster size ($S_{min}$) of 12. This process created the shaded regions overlaid on the clusters in Fig. \ref{fig:tSNE}, highlighting the major classes of band structures present in the dataset.

During the density-based clustering process, HDBSCAN assigns a probability score to each data point. This score indicates the likelihood of the data point (material) belonging to its assigned cluster. This probability effectively measures the range of density cut-off values for which a given data point remains part of its assigned cluster throughout the clustering process. By sorting the materials within each cluster based on their membership probability, it is possible to identify `exemplar' materials that are most representative of the properties characteristic of their respective clusters. These exemplar materials exhibit the highest membership probabilities within their clusters.  A full list of clusters with the materials ordered by membership probability can be found in \url{https://huggingface.co/datasets/2Dmatters/Elf_encoded_flat_band_materials/tree/main}. 

To visualize the separation and evolution of different band structure types across the t-SNE plot, we plotted band structure images corresponding to several exemplar materials from DBSCAN clusters alongside the t-SNE chart, Fig.~\ref{fig:tSNE}. The combination of the identified clusters and the t-SNE chart collectively form an electronic band structure genome for 2D flat-band materials, which serves as a comprehensive map of the diverse electronic properties found in these materials. This genome enables further exploration and a better understanding of the relationships among different flat-band materials based on their band structure characteristics. To further visualize the relationships among different types of band structures, we present a phylogenetic tree (Fig. \ref{fig:tSNE}a) that shows their hierarchical organisation. The leaves of the tree are opaque-coloured (same as in t-SNE) squares, the size of which shows the relative size of the clusters. Adjacent leaves belonging to the same DBSCAN group are further grouped in the tree shown in transparent shades of the same colours. We can see that clusters 18, 30-33, 26-27 and 14-15-17, which form isolated groups on the right in t-SNE, also form a separate branch in the tree. Clusters 38-49, 36-37, 28-29, 22-23 and 24-25 also form adjacent leaves in the tree highlighting their similar origin. 

\subsection{Analysis of global trends} \label{II.1}
The final stage of clustering using DBSCAN reveals distinct chemical patterns among the 2D flat-band materials. For example, clusters 26 and 27 predominantly contain halides, while clusters 24, 25, and 8 are rich in transition metal chalcogenides. Other notable chemical trends are highlighted on the t-SNE chart in Fig.~\ref{fig:tSNE}.

Interestingly, these chemical insights emerge solely from the unsupervised learning of electronic band structure features, without explicit input of the atomic composition. This can be attributed to the fact that materials containing elements from the same group in the periodic table often possess similar valence orbital structures, leading to comparable band arrangements and properties. Our convolutional autoencoder elf effectively captures these chemical patterns by learning the intrinsic similarities in the band structures.

\begin{figure}
\includegraphics[width=0.5\textwidth]{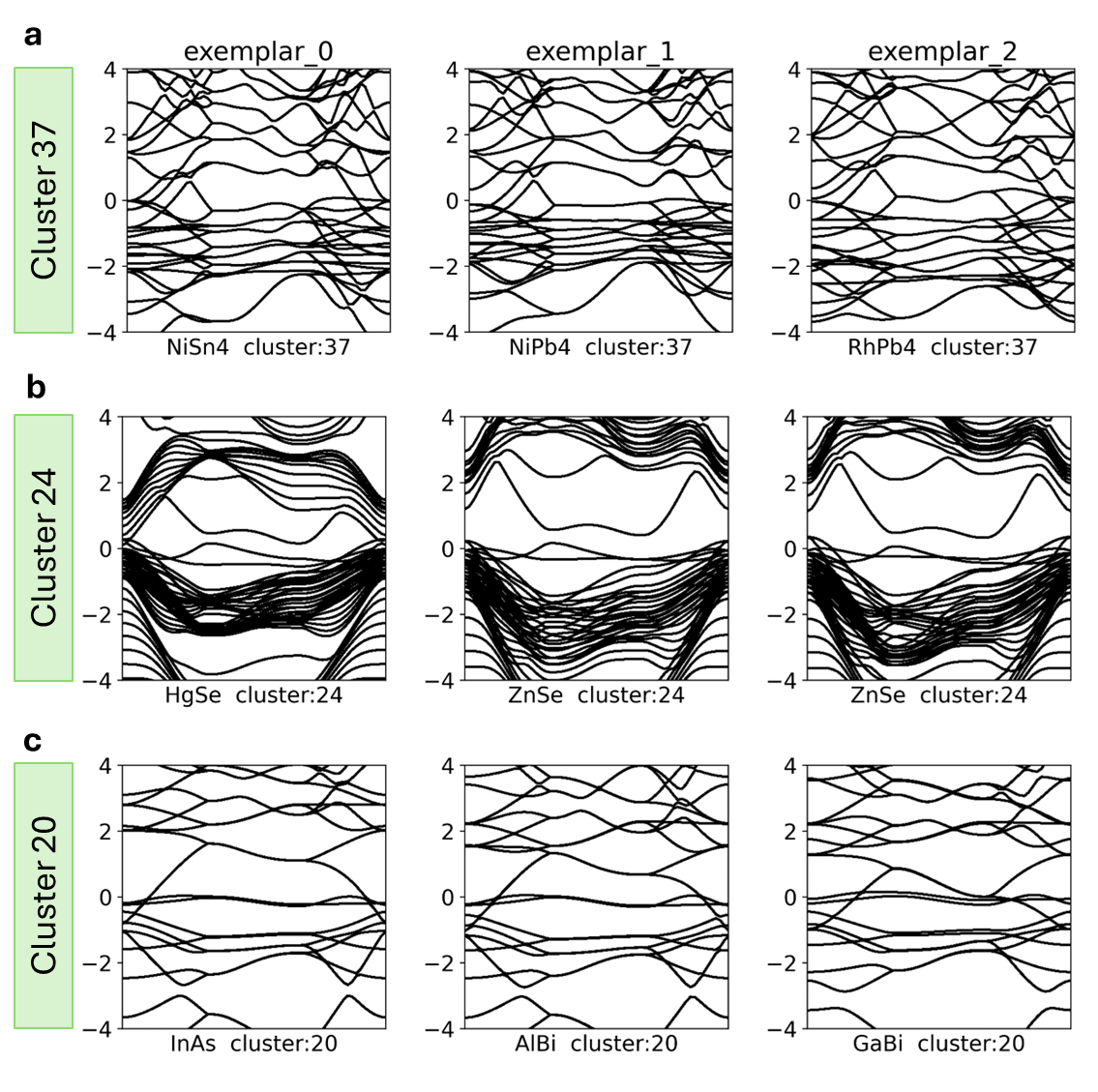}
\caption{\label{fig:clusters} Band structures of the top three exemplar materials of clusters 37 (\textbf{a}),  24 (\textbf{b}), and  20 (\textbf{c}).}
\end{figure}

However, the clustering is not solely determined by the chemical composition. The precise features of the band structure, such as band gaps, crossings, and gradients, act as additional classification constraints. Consequently, each cluster contains materials with similar band structure characteristics, despite potential variations in their chemical composition or crystal structure. This is exemplified in clusters 38-40 (Figure~\ref{fig:clusters}c, Dataset: \url{https://huggingface.co/datasets/2Dmatters/Elf_encoded_flat_band_materials}), where materials within each cluster exhibit characteristic band gaps, albeit deviations in the chemical composition. The convolutional nature of the autoencoder allows for a suitable margin of distortions and shifts in the electronic bands while still preserving the overall similarity. Materials with more pronounced deviations from the characteristic band structure of a cluster (exemplar band structure) are generally assigned lower membership probabilities.

The t-SNE chart in Figure~\ref{fig:tSNE} reveals a global partitioning of the flat-band materials based on their electronic band structure features. Notably, metallic, semi-metallic, semiconducting, and insulating states are well separated in the latent space. This clear partitioning validates the effectiveness of the elf-based fingerprinting approach in capturing meaningful electronic features. Furthermore, several clusters (e.g., 47-49 in Fig. \ref{fig:tSNE}d and 23-22-18 in Fig. \ref{fig:tSNE}g) are characterized by the presence of dense, flat bands near the Fermi energy. These clusters are particularly interesting from a materials discovery perspective, as they may host stronger electron-electron correlations and potentially exhibit exotic phenomena. The identification of these sub-groups demonstrates the power of the unsupervised learning approach in uncovering materials with desirable electronic properties.

\subsection{Local trends}  \label{II.2}
Within the individual clusters, HDBSCAN ensures a high degree of similarity among the band structure features of the clustered materials. Cluster 24, shown in Figure~\ref{fig:clusters}b, is a prime example of this. The cluster consists primarily of semiconductors, with Hg and Cd-based materials exhibiting additional complexity in their band structures. Depending on their specific structure, these materials can exhibit semimetallic behaviour \cite{Hg}, or a strain-tunable band gap, as observed in \ch{HgSe} and \ch{HgTe}, which can gradually transition into a topologically insulating phase \cite{Hg_2}). The unique band features near the Fermi level in cluster 24 suggest that these materials could potentially lead to a range of useful (opto)electronic applications. Notably, most of the materials in this cluster were predicted to be stable as 2D monolayers, making them promising candidates for van der Waals heterostructures \cite{Zn, Zn_2, Cd}.

One of the key advantages of our band structure-based fingerprinting approach is its ability to identify promising candidate materials that share electronic properties with well-studied compounds, even if their chemical compositions differ. By clustering materials based on their band structure similarity, we can flag computationally predicted compounds that have yet to be experimentally investigated but are likely to exhibit desirable properties.

For instance, cluster 46 contains 2dm-1072 \ch{Bi2O3}, a wide-band-gap semiconductor frequently used in heterostructures for its optoelectronic properties \cite{Bi1,Bi2}. Interestingly, the cluster also includes 2dm-3090 \ch{ZnMoO4} and 2dm-3226 \ch{Tl2SiSe3}, which displays a nearly identical band gap and distribution of flat bands but has not been previously studied. Moreover, the properties of two-dimensional \ch{ZnMoO4} and  \ch{Tl2SiSe3} have yet to be investigated in the literature. Based on its electronic similarity to \ch{Bi2O3}, they are flagged as promising candidates for similar optoelectronic applications, warranting further investigation. 

\begin{figure}
\includegraphics[width=\columnwidth]{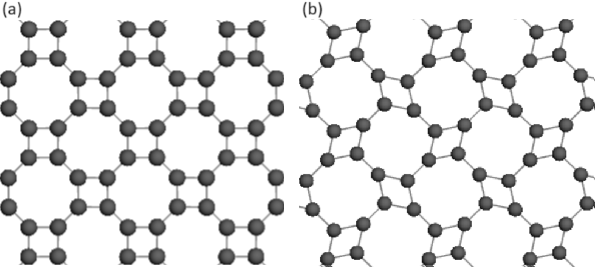}
\caption{\label{fig:square_octagonal} Schematics of square-octagonal lattice. (a) Symmetric square-octagon lattice structure. (b) Skewed square-octagon lattice structure.}
\end{figure}

Another example of this predictive power is demonstrated by cluster 20. This cluster is anchored by the well-known semiconductor \ch{InAs} (2dm-2474), which is considered one of the prime candidates for next-generation (opto)electronic devices due to its high mobility, large surface area, and direct band gap \cite{InAs}. The cluster also contains several less-studied materials, such as \ch{AlBi} (2dm-2252), a Rashba semiconductor \cite{AlBi}, and a group of thallium-based pnictides (2dm-2650 \ch{TlBi}, 2dm-2847 \ch{TlSb} and 2dm-2672 \ch{TlP}, all of which display strikingly similar band structures to InAs (Fig.~\ref{fig:clusters}c). Notably, these materials share a square-octagonal lattice (Fig.~\ref{fig:square_octagonal}), a structural motif known to host topological states \cite{Topol_2,Topol} and flat-band mediated correlated electron phenomena \cite{SqOct}. We predict that materials comprising cluster 20 will exhibit similarly promising electronic properties to \ch{InAs} and are worthy of further investigation. 

Interestingly, 2dm-2650 \ch{TlBi} and 2dm-2847 \ch{TlSb} in this cluster exhibit a unique asymmetrically skewed square-octagonal structure (Fig.~\ref{fig:square_octagonal}b), a lattice that has not been previously reported to host flat bands. Despite this structural distortion, these materials remain close to the other symmetric square-octagon lattice materials in our fingerprint space.  This underscores the ability of our approach to cluster materials with similar electronic properties even in the presence of structural variation. 

This structural flexibility extends to the clustering of materials with entirely different crystal structures based on their common band features. For example, clusters 46-49 contain a variety of wide band gap ($\approx$3eV) flat-band semiconductors with different crystal structures. Notably, cluster 46 includes 2dm-3090 \ch{ZnMoO4}, which exhibits a unique edge-sharing zigzag octahedra chain sublattice, a material whose flat-band formation mechanisms are yet to be understood.  

While some of the observed flat-band clusters can be readily explained by well-known flat-band physics, such as the localisation of electron wavefunctions in the f-orbitals of lanthanides and actinides \cite{flat_Cat}, many others emerge beyond the known flat-band paradigm. For instance, there are bilayer flat-band structures commonly consisting of two stacked monolayers that individually exhibit flat bands, with well-studied examples including stacked square and Kagome arrangements. However, stacked centred-orthorhombic-square chains have only been reported very recently \cite{Anupam}. 

By lowering the minimum cluster size to $N_{c}=4$, we identify a cluster of four chemically similar materials: \ch{NdF3} (2dm-321), \ch{TbF3} (2dm-441), \ch{YF3} (2dm-553), and \ch{SmF3} (2dm-875). All of these compounds exhibit a bilayer sublattice structure composed of stacked centred-orthorhombic-square chains and possess flat electronic bands. The mechanisms leading to the emergence of flat bands in this type of lattice have yet to be uncovered.

Cluster 37, shown in Fig.~\ref{fig:clusters}a, contains a mixture of group 8-9 transition metals with group 14 elements (Si/Pb/Sn) with the $AB_{4}$ and $AB_{3}$ stoichiometries. The two groups of materials with these stoichiometries exhibit distinct tetragonal structures with different point groups (4/mm and 422, respectively). Further inspection of the orbitally projected band structures for these materials reveals that the nearly plane-flat bands arise from mixtures between the A and B elements, whose element groups are shared for both stoichiometries. This suggests that the plane-flat bands in this cluster likely result from a non-trivial interplay between the shared chemistry of these materials and the tetragonal $D_{4}$ abstract symmetry group (shared by both 4/mm and 422 point groups) under which both lattices fall. More investigation is required to understand this interplay in detail, adding to the known non-trivial mechanisms that result in flat electronic bands beyond the traditional picture of orbital overlap within a single element sublattice \cite{flat_band_overlap}.

\subsection{Comparison to structure fingerprints}  \label{II.3}
While our auto-encoded fingerprint is highly effective at clustering materials based on their electronic properties, it is not completely robust to structural distortions. Small shifts in a lattice can alter the high-symmetry points in reciprocal space, which in turn affect the band structure image used as input to the network. However, the likelihood of this occurring in practice is minimal. We further inspected the clusters (see \url{https://huggingface.co/datasets/2Dmatters/Elf_encoded_flat_band_materials/tree/main}), and found that the ability to extend beyond structural similarity and cluster materials based on their emergent electronic features is a general capability of our approach.

\begin{figure*}
\includegraphics[width=0.8\textwidth]{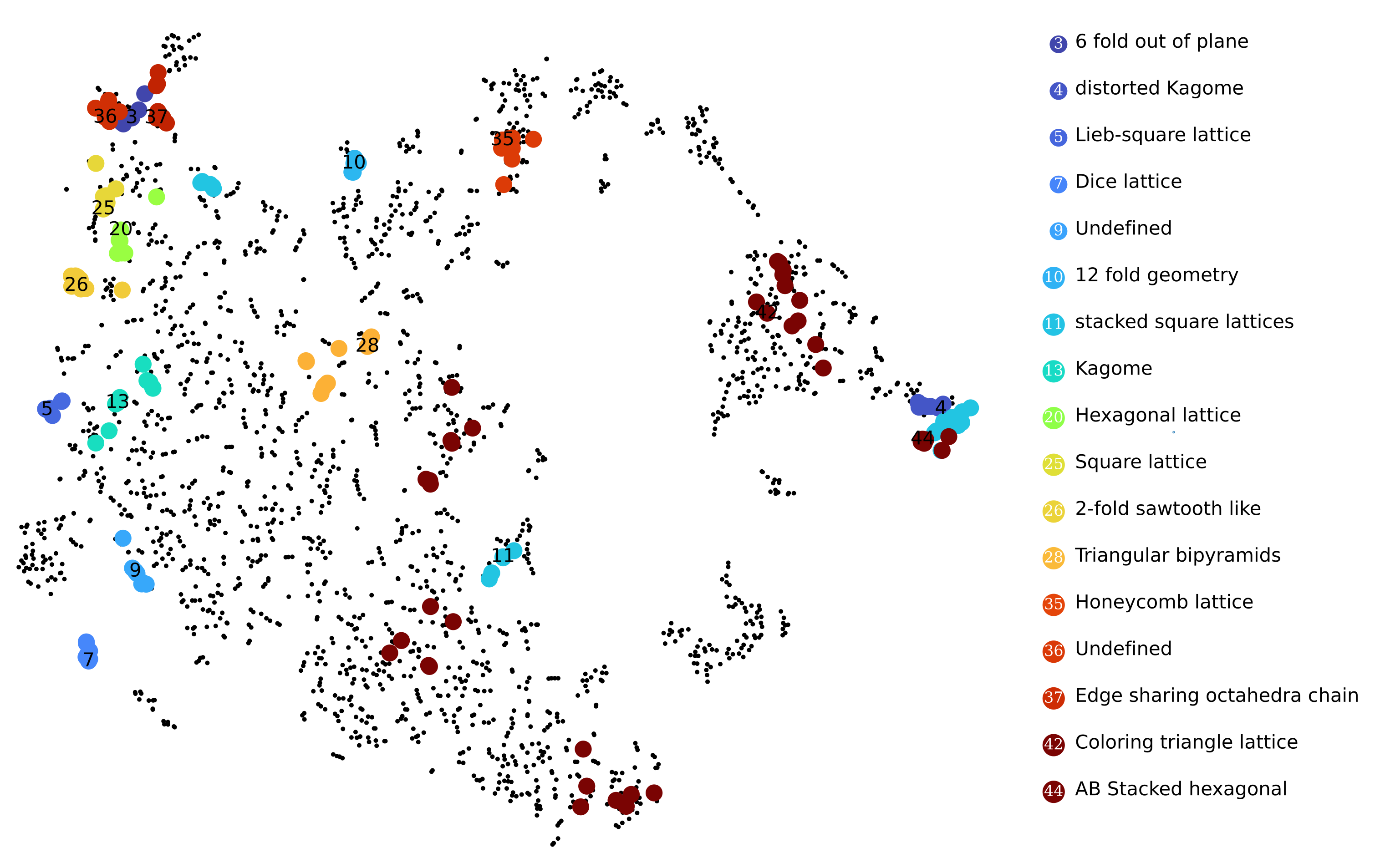} 
\caption{\label{fig:str_fing} Clusters of structural fingerprint marked in the electronic fingerprint space to visualise the extent to which electronic fingerprints also cluster similar structures together.}
\end{figure*}

To directly compare our elf fingerprints to structure-based fingerprints, we generated structural fingerprints for the flat band sublattices following our previous work \cite{Anupam} for each of 2127 flat-band materials studied using CrystalNNFingerprint (CNNF) \cite{Crystal}. We then clustered these CNNF fingerprints using HDBSCAN, which resulted in a total of 45 clusters. To visualize the relationship between structural (CNNF) and electronic fingerprints (elf), we assigned each material in the elf t-SNE space its corresponding CNNF cluster label. This allows us to identify any localized groupings of CNNF clusters within the electronic fingerprint space, thereby revealing correlations between structural and band-structure similarity. The resulting plot is shown in Fig. \ref{fig:str_fing}, with the identified structural motifs within each cluster listed on the right. 

We find that out of the 45 CNNF clusters, 17 form localized groups of more than 4 materials within the electronic fingerprint space. This clearly demonstrates that certain structural motifs tend to give rise to similar band structures. For example, in Fig. \ref{fig:str_fing}, cluster 13 predominantly contains Kagome sublattices, cluster 7 hosts dice lattices, cluster 35 is composed of honeycomb lattices, and cluster 25 features square lattices. These lattice-specific groups form fairly isolated islands in the elf space. 

However, it is important to note that the number of CNNF clusters forming distinct groups in electronic fingerprint space is much lower than the total number of CNNF clusters. This implies that many materials sharing similar sublattice structures can indeed exhibit very different band structures. This can be attributed to the variations in the orbital composition of the materials, as well as differences in the atomic sizes and electronegativities of the constituent elements, which also play a role in determining the electronic distribution around the structural motifs.

\subsection{Duplicate detection}  \label{II.4}
In general, when 2dmatpedia generates ``bottom-up'' materials via element substitution, the structure is allowed to relax to equilibrium bond lengths and angles without changing the crystal symmetries. Conversely, based on 2Dmatpedia's ``top-down'' generation mechanism, we should expect that some layers exfoliated from unique 3D structures in the Materials Project will be equivalent in two dimensions. Subsequently, this could generate identical materials when the bottom-up element substitution chains intersect such that the constituent elements are also the same. These are duplicate materials, and most would have been removed by 2Dmatpedia using structure-matching tools available from the pymatgen library \cite{pymatgen}.

Our algorithm enables the automatic detection of such duplicates, as demonstrated in Fig. \ref{fig:clusters}b. In this example, we see two entries of \ch{ZnSe} in cluster 24 exhibiting nearly identical band structures.   Our elf detected several other pairs of duplicate materials with similar band structures, differing only by small structural distortions. The 2dmatpedia IDs of these materials are listed in Table 1, with each duplicate pair sharing the same chemical formula. Using this approach, we found up to 40 potential duplicate entries (listed in \url{https://huggingface.co/datasets/2Dmatters/Elf_encoded_flat_band_materials/tree/main}). However, further investigation is necessary to determine the stability of these materials within the accuracy of DFT calculations.

We recommend verifying the comparative stabilities of the identified duplicate pairs and removing the less stable entries from the 2Dmatpedia database. Our fingerprint, based solely on electronic band structures, enables the identification of fundamentally equivalent materials that differ only by small structural distortions, setting it apart from structure-based methods. This capability is particularly valuable for maintaining accurate and concise materials databases, which is a prerequisite for high-throughput computational materials discovery.

\begin{table}[]
    \centering
\begin{tabular}{ |c|c|c|c| }
\hline
 In$_{2}$S & Tl$_{2}$Te & PI$_{3}$ & Te$_{2}$Se  \\
 \hline
 2dm-1845 & 2dm-1554 & 2dm-495 & 2dm-1823  \\ 
 2dm-1986 & 2dm-1521 & 2dm-2009 & 2dm-1596 \\
\hline
\end{tabular}

\begin{tabular}{ |c|c|c|c| }
\hline
 SrLaCl$_{5}$ & AsBr$_{3}$ & I$_{3}$N & ZnSe  \\
 \hline
  2dm-5260 & 2dm-2624 & 2dm-2010 & 2dm-2113 \\ 
  2dm-5422 & 2dm-4881 & 2dm-726 & 2dm-2321\\
\hline
\end{tabular}

    \caption{Examples of the duplicates identified by elf.}
    \label{tab:1}
\end{table}

\section{Conclusions}  \label{III}
In this work, we have proposed a novel material fingerprinting method based on electronic band structures and demonstrated its advantages over structure-based methods, complementing existing techniques employed for material similarity search. We applied our fingerprinting and clustering framework (elf) to two-dimensional materials exhibiting flat bands to determine chemical and electronic property trends, elucidating multiple chemical and structure groups for further investigation. This fully unsupervised approach is a stepping stone in the realisation of the autonomous materials discovery paradigm. 

Similarity search in material properties has become one of the main challenges in modern materials science. Very recently, Google Deepmind significantly increased the number of known stable crystals with GNoME (Graph Neural Networks for Materials Exploration) \cite{DEEP_1}, releasing an unprecedented number of candidate materials. In this work, we have demonstrated for the first time that material fingerprints deep-learned from electronic band-structure features prove to be a robust tool in linking computationally generated materials to already synthesised compounds exhibiting important emergent properties. This will help widen the bottleneck between material prediction and synthesis of the most promising candidate materials, a crucial task as we move into the new paradigm of AI-driven materials discovery.

In future work, we plan to apply our approach to larger databases, such as the Materials Project, containing 3D materials, and investigate different electronic phenomena emerging from non-trivial band features such as high $T_{c}$ superconductivity and novel topological phases.

\section{Methods}  \label{M}
\subsection{ Network Training and Fingerprint} \label{M.1}
Previous studies have employed simple fingerprint vectors directly extracted from the electronic band structure of a material \cite{E_2,E_1} to cluster materials' electronic properties. However, when applying similar techniques to the subset of flat-band materials from 2Dmatpedia, we found that the results were dominated by noise, with materials relatively uniformly spread across fingerprint space. This issue mainly arises from the reliance of those techniques on integrated variables of all electronic bands in some energy range, like the density of states (DoS). As a result, materials sharing meaningful electronic properties may be far apart in fingerprint space if they happen to have a different number of bands passing through some energy range.

To address this issue, we have proposed a fingerprint, based solely on the electronic band structure features of a material. The autoencoder we trained to encode the band feature fingerprints is based on the ResNet series of convolutional neural networks\cite{Res_11, Res_22}, which have found extensive applications in feature extraction problems across the medical and physical sciences \cite{Res_1,Res_2}. We used the first 16 layers of ResNet18 as an encoder and `transposed' it by replacing convolutions with deconvolutions to obtain the decoder. The resulting network is fully convolutional and features skip connections and batch normalization layers, which are characteristic of ResNet models and enable the deep model to converge. 

To train the network, we plotted band structure data within a $\pm$4 eV range of the Fermi energy, binarized the plots, and resized them as 224x224 pixel images. Limiting the energy range to this region around the Fermi energy focuses on the crucial features of the band structure, preventing excessive uniqueness in materials that could lead to reduced clustering power.

The input to the network is a 224x224 matrix of zeros and ones, representing the pixels of the band structure image. The network takes 3 layers of size 224x224 to represent the RGB colour values of the pixels but these are redundant for our black and white images. The network then predicts an output matrix of the same size as the input. When this output matrix is plotted, we obtain a prediction of the input band structure, based only on the information from the compressed latent layer representation in the centre of the network. 

During the optimization process, the network is trained to minimise the binary cross-entropy (BCE) loss \cite{BCE} between input and output images. To generate an accurate prediction of the input image from the much smaller set of numbers in the latent representation, the network is forced to encode compressed features of the band structure image in this latent space. Additionally, during training, we applied random noise to regions of the input images with a probability of 0.5. 

To achieve a balance between the flexibility needed for high reconstruction accuracy and the dimensionality reduction required for improved clustering, we chose a flattened length of 98 for the latent space. The latent space size was set to a 7x7 matrix with two parallel channels. With this network architecture and the application of noise during training, we observed the training loss stabilise at 0.282 (using BCE) after 30 epochs. Performance of the network on the validation set remained within 5$\%$ of the training loss throughout the training process. 

We can interpret the physical features learned by the network by inspecting its latent space representations. Due to the purely convolutional architecture of the network, we expect soft correlation between specific regions of the input image and specific regions of the latent space matrix. To visualise this, we run the band structure of 2dm-1's (IrF$_2$) through the encoder, and systematically varied the value of one dimension of the resulting encoded representation by a $\Delta$. The full set of slightly altered latent space values is then decoded, and any effect of the change will be observed in the features of the reconstructed image.

\begin{figure*}
\includegraphics[width=\textwidth]{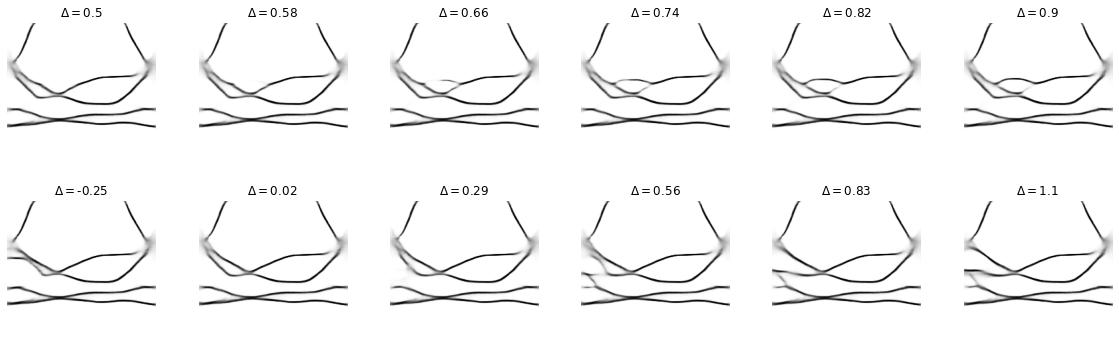}
\caption{ Reconstructed band structures showing the effect of changing one dimension of 2dm-1's latent space. The first and second rows correspond to changes in the latent dimensions [2,2] and [3,0] respectively, of channel 2. The ranges of $\Delta$ have been chosen to display a single altering feature. By varying more, different possible splittings can emerge in the same region until distortion of the band structure for extreme values $|\Delta|>2$ which pushes outside the range of latent space values for any encoded two-dimensional material.}
\label{fig:reconstructions}
\end{figure*}

We changed the latent dimension of channel 2 at matrix position (2,2) by a value $\Delta$ in the range 0.5 to 0.9, and the dimension at matrix position (3,0) was changed by $\Delta$ in the range of 0.25 to 1.1. The resulting reconstructions are displayed in Figure ~\ref{fig:reconstructions}, with $\Delta$=0 corresponding to the material's original band structure.

We observe that, because of the learned features, the auto-encoder can generalise to generate entirely sensible band structures of materials that, in theory, do not exist, by simple manipulation of the latent space. This helps to elucidate the meanings of the individual latent space dimensions. Moreover, due to the compression, there is generally overlap in the latent space regions, and this overlap can help obtain band structures of two seemingly different material groups by continuously tuning some parameters of the latent vector. However, that exercise is out of the scope of this article.

\subsection{ Multi-stage clustering module} \label{M.2}
To classify the machine-learned fingerprints, we employed a completely unsupervised multi-stage algorithm. HDBSCAN \cite{HDB} was first used to discern regions of high density in the fingerprint space and suggest a hierarchical structure for these clusters. This serves as a stringent identifier of band structure feature similarity among the materials. 
HDBSCAN, being density-based, facilitates much more general cluster shapes compared to the common ellipsoid-based k-means method. Additionally, it allows us to obtain hierarchical cluster information without being as sensitive to noise in the data (from structural distortions).

To offer a complementary and independent view of our 98-dimensional fingerprint space, t-distributed Stochastic Neighbor Embedding (t-SNE) was used. t-SNE excels at dimensional reduction while preserving the local and global distance relations of points\cite{t-sne}. This sets it apart from other approaches such as Locally Linear Embedding (LLE) \cite{LLE}, Hessian Eigenmaps \cite{Hessian}, and UMAP \cite{UMAP}. In our analysis, we did not observe any measurable improvement in our results when using these alternative methods compared to t-SNE. 
Furthermore, to visualize relations between different clusters, we used another clustering technique DBSCAN, which allows even arbitrary-shaped connections to create larger groups. 

The Minkowski distance with exponent p=0.2 was used during the clustering process, as this metric is known to scale better than Euclidean (p=2) and Manhattan (p=1) metrics to high dimensional vector spaces \cite{Metric}. 

\begin{figure*}
\includegraphics[width=\textwidth]{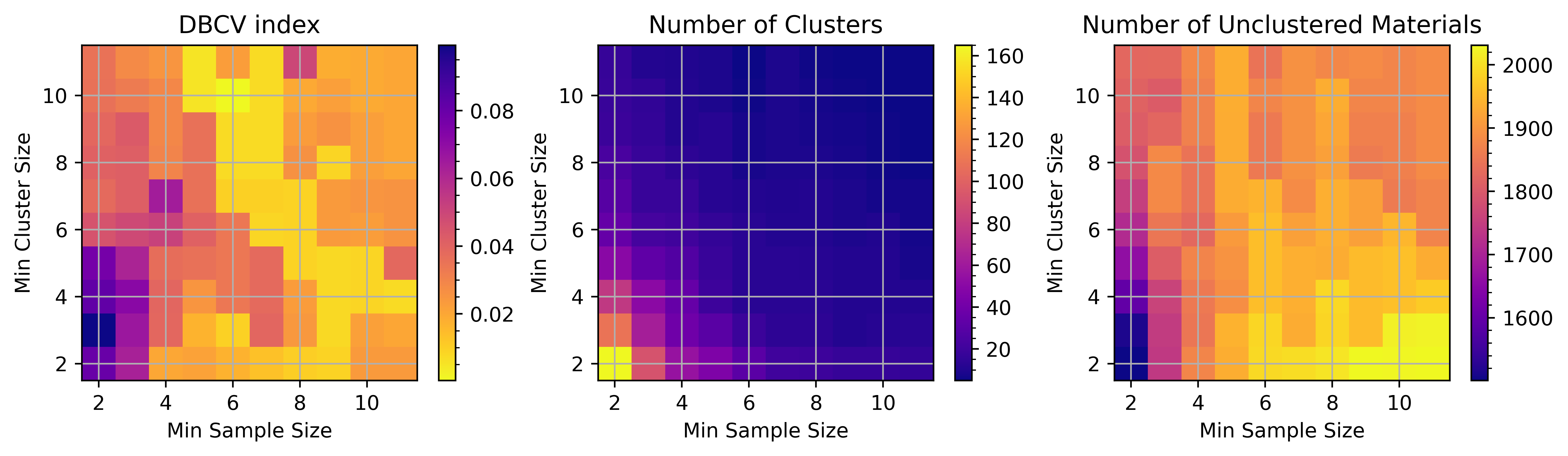} 
\caption{\label{fig:opt} Metrics for assessing the validity of a clustering solution as a function of HDBSCAN's minimum cluster size and minimum sample size variables.}
\end{figure*}

The two primary free parameters of HDBSCAN (minimum cluster size, $N_{c}$, and minimum sample size, $N_{s}$), were optimised by considering their effect on three metrics quantifying the quality of the resulting clustering solution. These were the number of clusters formed, the number of unclassified materials, and the `density based clustering validation' (DBCV) index \cite{DBCV}. The DBCV index evaluates the compactness of a clustering solution by comparing the sparseness of clusters (based on the point in the cluster with the largest core distance measure) with the inter-cluster separation. Thus, a higher value indicates more compact, well-separated clusters and an overall better clustering solution.

$N_{s}$ and $N_{c}$ were both varied from 2 to 12, and the metrics above were calculated for the resulting clustering solution. These are displayed as colour maps in Figure~\ref{fig:opt}. Initially, the number of unclassified materials increases, indicating the presence of many difficult-to-classify materials that get forced out of clusters as the clustering parameters become more stringent, requiring larger and more compact clusters. These behaviours are typical of material clustering solutions \cite{Anupam} using this approach. Considering these factors, we chose $N_{c}$=5 and $N_{s}$=2. This achieves a relatively large DBCV index while minimising the number of unclassified materials and keeping the number of clusters bounded enough to effectively represent the major band structure groups among flat-band materials.

For t-SNE, a perplexity of 10 was found to give the optimal 2D representation and general agreement with the HDBSCAN clusters. Finally, for DBSCAN, the parameter $\epsilon =25$ was chosen which allowed identifying visibly separate regions in the t-SNE projection.

\section*{Data availability}
All relevant data are available at our HuggingFace repository: \href{https://huggingface.co/datasets/2Dmatters/Elf_encoded_flat_band_materials}{Elf\_encoded\_flat\_band\_materials}.

\section*{Code availability}
The code used to generate the results discussed in this work is available from our public GitHub repository: \href{https://github.com/ThomasWarford/band-fingerprint}{Band-Fingerprinting-Repository} under the `2dmatpedia' branch.

\section*{Acknowledgements}
This research was supported by the European Research Council (ERC) under the European Union's Horizon 2020 research and innovation program (Grant Agreement No. 865590) and the Research Council UK [BB/X003736/1]. Q.Y. acknowledges the funding from Royal Society University Research Fellowship URF$\backslash$R1$\backslash$221096 and UK Research and Innovation Grant [EP/X017575/1].

\section*{Author contributions}
H.P. and T.W. equally contributed in carrying out the calculations, programming and analysis of the data used in this work. This includes the conception of the novel fingerprinting algorithm and the idea of using a convolutional autoencoder. A.M. and A.B. conceived the research plan to investigate the potential of electronic band-based fingerprints for understanding 2D flat-band materials. They also introduced the idea of using HDBSCAN and t-SNE as analysis methods. All authors participated in discussions and contributed to writing the manuscript. 

\section*{Competing Interests}
The authors declare no competing interests.

\bibliography{references}

\begin{thebibliography}{55}%
\makeatletter
\providecommand \@ifxundefined [1]{%
 \@ifx{#1\undefined}
}%
\providecommand \@ifnum [1]{%
 \ifnum #1\expandafter \@firstoftwo
 \else \expandafter \@secondoftwo
 \fi
}%
\providecommand \@ifx [1]{%
 \ifx #1\expandafter \@firstoftwo
 \else \expandafter \@secondoftwo
 \fi
}%
\providecommand \natexlab [1]{#1}%
\providecommand \enquote  [1]{``#1''}%
\providecommand \bibnamefont  [1]{#1}%
\providecommand \bibfnamefont [1]{#1}%
\providecommand \citenamefont [1]{#1}%
\providecommand \href@noop [0]{\@secondoftwo}%
\providecommand \href [0]{\begingroup \@sanitize@url \@href}%
\providecommand \@href[1]{\@@startlink{#1}\@@href}%
\providecommand \@@href[1]{\endgroup#1\@@endlink}%
\providecommand \@sanitize@url [0]{\catcode `\\12\catcode `\$12\catcode `\&12\catcode `\#12\catcode `\^12\catcode `\_12\catcode `\%12\relax}%
\providecommand \@@startlink[1]{}%
\providecommand \@@endlink[0]{}%
\providecommand \url  [0]{\begingroup\@sanitize@url \@url }%
\providecommand \@url [1]{\endgroup\@href {#1}{\urlprefix }}%
\providecommand \urlprefix  [0]{URL }%
\providecommand \Eprint [0]{\href }%
\providecommand \doibase [0]{https://doi.org/}%
\providecommand \selectlanguage [0]{\@gobble}%
\providecommand \bibinfo  [0]{\@secondoftwo}%
\providecommand \bibfield  [0]{\@secondoftwo}%
\providecommand \translation [1]{[#1]}%
\providecommand \BibitemOpen [0]{}%
\providecommand \bibitemStop [0]{}%
\providecommand \bibitemNoStop [0]{.\EOS\space}%
\providecommand \EOS [0]{\spacefactor3000\relax}%
\providecommand \BibitemShut  [1]{\csname bibitem#1\endcsname}%
\let\auto@bib@innerbib\@empty
%</preamble>
\bibitem [{\citenamefont {Merchant}\ \emph {et~al.}(2023)\citenamefont {Merchant}, \citenamefont {Batzner}, \citenamefont {Schoenholz}, \citenamefont {Aykol}, \citenamefont {Cheon},\ and\ \citenamefont {Cubuk}}]{DEEP_1}%
  \BibitemOpen
  \bibfield  {author} {\bibinfo {author} {\bibfnamefont {A.}~\bibnamefont {Merchant}}, \bibinfo {author} {\bibfnamefont {S.}~\bibnamefont {Batzner}}, \bibinfo {author} {\bibfnamefont {S.~S.}\ \bibnamefont {Schoenholz}}, \bibinfo {author} {\bibfnamefont {M.}~\bibnamefont {Aykol}}, \bibinfo {author} {\bibfnamefont {G.}~\bibnamefont {Cheon}},\ and\ \bibinfo {author} {\bibfnamefont {E.~D.}\ \bibnamefont {Cubuk}},\ }\bibfield  {title} {\bibinfo {title} {Scaling deep learning for materials discovery},\ }\href@noop {} {\bibfield  {journal} {\bibinfo  {journal} {Nature}\ }\textbf {\bibinfo {volume} {624}},\ \bibinfo {pages} {80} (\bibinfo {year} {2023})}\BibitemShut {NoStop}%
\bibitem [{\citenamefont {Rashid}\ \emph {et~al.}(2024)\citenamefont {Rashid}, \citenamefont {Lazarev}, \citenamefont {Kazeev}, \citenamefont {Novoselov},\ and\ \citenamefont {Ustyuzhanin}}]{rashid2024review}%
  \BibitemOpen
  \bibfield  {author} {\bibinfo {author} {\bibfnamefont {A.}~\bibnamefont {Rashid}}, \bibinfo {author} {\bibfnamefont {M.}~\bibnamefont {Lazarev}}, \bibinfo {author} {\bibfnamefont {N.}~\bibnamefont {Kazeev}}, \bibinfo {author} {\bibfnamefont {K.}~\bibnamefont {Novoselov}},\ and\ \bibinfo {author} {\bibfnamefont {A.}~\bibnamefont {Ustyuzhanin}},\ }\bibfield  {title} {\bibinfo {title} {Review on automated {2D} material design},\ }\href@noop {} {\bibfield  {journal} {\bibinfo  {journal} {2D Materials}\ } (\bibinfo {year} {2024})}\BibitemShut {NoStop}%
\bibitem [{\citenamefont {Leeman}\ \emph {et~al.}(2024)\citenamefont {Leeman}, \citenamefont {Liu}, \citenamefont {Stiles}, \citenamefont {Lee}, \citenamefont {Bhatt}, \citenamefont {Schoop},\ and\ \citenamefont {Palgrave}}]{leeman2024challenges}%
  \BibitemOpen
  \bibfield  {author} {\bibinfo {author} {\bibfnamefont {J.}~\bibnamefont {Leeman}}, \bibinfo {author} {\bibfnamefont {Y.}~\bibnamefont {Liu}}, \bibinfo {author} {\bibfnamefont {J.}~\bibnamefont {Stiles}}, \bibinfo {author} {\bibfnamefont {S.~B.}\ \bibnamefont {Lee}}, \bibinfo {author} {\bibfnamefont {P.}~\bibnamefont {Bhatt}}, \bibinfo {author} {\bibfnamefont {L.~M.}\ \bibnamefont {Schoop}},\ and\ \bibinfo {author} {\bibfnamefont {R.~G.}\ \bibnamefont {Palgrave}},\ }\bibfield  {title} {\bibinfo {title} {Challenges in high-throughput inorganic materials prediction and autonomous synthesis},\ }\href@noop {} {\bibfield  {journal} {\bibinfo  {journal} {PRX Energy}\ }\textbf {\bibinfo {volume} {3}},\ \bibinfo {pages} {011002} (\bibinfo {year} {2024})}\BibitemShut {NoStop}%
\bibitem [{\citenamefont {Jain}(2024)}]{jain2024machine}%
  \BibitemOpen
  \bibfield  {author} {\bibinfo {author} {\bibfnamefont {A.}~\bibnamefont {Jain}},\ }\bibfield  {title} {\bibinfo {title} {Machine learning in materials research: developments over the last decade and challenges for the future},\ }\href@noop {} {\bibfield  {journal} {\bibinfo  {journal} {ChemRxiv: 10.26434}\ } (\bibinfo {year} {2024})}\BibitemShut {NoStop}%
\bibitem [{\citenamefont {Lyngby}\ and\ \citenamefont {Thygesen}(2024)}]{lyngby2024ab}%
  \BibitemOpen
  \bibfield  {author} {\bibinfo {author} {\bibfnamefont {P.}~\bibnamefont {Lyngby}}\ and\ \bibinfo {author} {\bibfnamefont {K.~S.}\ \bibnamefont {Thygesen}},\ }\bibfield  {title} {\bibinfo {title} {Ab initio property characterisation of thousands of previously {unknown 2D} materials},\ }\href@noop {} {\bibfield  {journal} {\bibinfo  {journal} {arXiv: 2402.02783}\ } (\bibinfo {year} {2024})}\BibitemShut {NoStop}%
\bibitem [{\citenamefont {Cheetham}\ and\ \citenamefont {Seshadri}(2024)}]{cheetham2024artificial}%
  \BibitemOpen
  \bibfield  {author} {\bibinfo {author} {\bibfnamefont {A.~K.}\ \bibnamefont {Cheetham}}\ and\ \bibinfo {author} {\bibfnamefont {R.}~\bibnamefont {Seshadri}},\ }\bibfield  {title} {\bibinfo {title} {{Artificial Intelligence Driving Materials Discovery? Perspective on the Article: Scaling Deep Learning for Materials Discovery}},\ }\href@noop {} {\bibfield  {journal} {\bibinfo  {journal} {Chemistry of Materials}\ }\textbf {\bibinfo {volume} {36}},\ \bibinfo {pages} {3490} (\bibinfo {year} {2024})}\BibitemShut {NoStop}%
\bibitem [{\citenamefont {Alverson}\ \emph {et~al.}(2024)\citenamefont {Alverson}, \citenamefont {Baird}, \citenamefont {Murdock}, \citenamefont {Johnson}, \citenamefont {Sparks} \emph {et~al.}}]{alverson2024generative}%
  \BibitemOpen
  \bibfield  {author} {\bibinfo {author} {\bibfnamefont {M.}~\bibnamefont {Alverson}}, \bibinfo {author} {\bibfnamefont {S.~G.}\ \bibnamefont {Baird}}, \bibinfo {author} {\bibfnamefont {R.}~\bibnamefont {Murdock}}, \bibinfo {author} {\bibfnamefont {J.}~\bibnamefont {Johnson}}, \bibinfo {author} {\bibfnamefont {T.~D.}\ \bibnamefont {Sparks}}, \emph {et~al.},\ }\bibfield  {title} {\bibinfo {title} {Generative adversarial networks and diffusion models in material discovery},\ }\href@noop {} {\bibfield  {journal} {\bibinfo  {journal} {Digital Discovery}\ }\textbf {\bibinfo {volume} {3}},\ \bibinfo {pages} {62} (\bibinfo {year} {2024})}\BibitemShut {NoStop}%
\bibitem [{\citenamefont {Adam}(2024)}]{adam2024automated}%
  \BibitemOpen
  \bibfield  {author} {\bibinfo {author} {\bibfnamefont {D.}~\bibnamefont {Adam}},\ }\bibfield  {title} {\bibinfo {title} {The automated lab of tomorrow},\ }\href@noop {} {\bibfield  {journal} {\bibinfo  {journal} {Proceedings of the National Academy of Sciences}\ }\textbf {\bibinfo {volume} {121}},\ \bibinfo {pages} {e2406320121} (\bibinfo {year} {2024})}\BibitemShut {NoStop}%
\bibitem [{\citenamefont {Wang}\ \emph {et~al.}(2023)\citenamefont {Wang}, \citenamefont {Fu}, \citenamefont {Du}, \citenamefont {Gao}, \citenamefont {Huang}, \citenamefont {Liu}, \citenamefont {Chandak}, \citenamefont {Liu}, \citenamefont {Van~Katwyk}, \citenamefont {Deac} \emph {et~al.}}]{wang2023scientific}%
  \BibitemOpen
  \bibfield  {author} {\bibinfo {author} {\bibfnamefont {H.}~\bibnamefont {Wang}}, \bibinfo {author} {\bibfnamefont {T.}~\bibnamefont {Fu}}, \bibinfo {author} {\bibfnamefont {Y.}~\bibnamefont {Du}}, \bibinfo {author} {\bibfnamefont {W.}~\bibnamefont {Gao}}, \bibinfo {author} {\bibfnamefont {K.}~\bibnamefont {Huang}}, \bibinfo {author} {\bibfnamefont {Z.}~\bibnamefont {Liu}}, \bibinfo {author} {\bibfnamefont {P.}~\bibnamefont {Chandak}}, \bibinfo {author} {\bibfnamefont {S.}~\bibnamefont {Liu}}, \bibinfo {author} {\bibfnamefont {P.}~\bibnamefont {Van~Katwyk}}, \bibinfo {author} {\bibfnamefont {A.}~\bibnamefont {Deac}}, \emph {et~al.},\ }\bibfield  {title} {\bibinfo {title} {Scientific discovery in the age of artificial intelligence},\ }\href@noop {} {\bibfield  {journal} {\bibinfo  {journal} {Nature}\ }\textbf {\bibinfo {volume} {620}},\ \bibinfo {pages} {47} (\bibinfo {year} {2023})}\BibitemShut {NoStop}%
\bibitem [{\citenamefont {Szymanski}\ \emph {et~al.}(2023)\citenamefont {Szymanski}, \citenamefont {Rendy}, \citenamefont {Fei}, \citenamefont {Kumar}, \citenamefont {He}, \citenamefont {Milsted}, \citenamefont {McDermott}, \citenamefont {Gallant}, \citenamefont {Cubuk}, \citenamefont {Merchant} \emph {et~al.}}]{szymanski2023autonomous}%
  \BibitemOpen
  \bibfield  {author} {\bibinfo {author} {\bibfnamefont {N.~J.}\ \bibnamefont {Szymanski}}, \bibinfo {author} {\bibfnamefont {B.}~\bibnamefont {Rendy}}, \bibinfo {author} {\bibfnamefont {Y.}~\bibnamefont {Fei}}, \bibinfo {author} {\bibfnamefont {R.~E.}\ \bibnamefont {Kumar}}, \bibinfo {author} {\bibfnamefont {T.}~\bibnamefont {He}}, \bibinfo {author} {\bibfnamefont {D.}~\bibnamefont {Milsted}}, \bibinfo {author} {\bibfnamefont {M.~J.}\ \bibnamefont {McDermott}}, \bibinfo {author} {\bibfnamefont {M.}~\bibnamefont {Gallant}}, \bibinfo {author} {\bibfnamefont {E.~D.}\ \bibnamefont {Cubuk}}, \bibinfo {author} {\bibfnamefont {A.}~\bibnamefont {Merchant}}, \emph {et~al.},\ }\bibfield  {title} {\bibinfo {title} {An autonomous laboratory for the accelerated synthesis of novel materials},\ }\href@noop {} {\bibfield  {journal} {\bibinfo  {journal} {Nature}\ }\textbf {\bibinfo {volume} {624}},\ \bibinfo {pages} {86} (\bibinfo {year} {2023})}\BibitemShut {NoStop}%
\bibitem [{\citenamefont {Bhattacharya}\ \emph {et~al.}(2023)\citenamefont {Bhattacharya}, \citenamefont {Timokhin}, \citenamefont {Chatterjee}, \citenamefont {Yang},\ and\ \citenamefont {Mishchenko}}]{Anupam}%
  \BibitemOpen
  \bibfield  {author} {\bibinfo {author} {\bibfnamefont {A.}~\bibnamefont {Bhattacharya}}, \bibinfo {author} {\bibfnamefont {I.}~\bibnamefont {Timokhin}}, \bibinfo {author} {\bibfnamefont {R.}~\bibnamefont {Chatterjee}}, \bibinfo {author} {\bibfnamefont {Q.}~\bibnamefont {Yang}},\ and\ \bibinfo {author} {\bibfnamefont {A.}~\bibnamefont {Mishchenko}},\ }\bibfield  {title} {\bibinfo {title} {Deep learning approach to genome of two-dimensional materials with flat electronic bands},\ }\href@noop {} {\bibfield  {journal} {\bibinfo  {journal} {npj Computational Materials}\ }\textbf {\bibinfo {volume} {9}},\ \bibinfo {pages} {101} (\bibinfo {year} {2023})}\BibitemShut {NoStop}%
\bibitem [{\citenamefont {Nunez}(2019)}]{feature_2}%
  \BibitemOpen
  \bibfield  {author} {\bibinfo {author} {\bibfnamefont {M.}~\bibnamefont {Nunez}},\ }\bibfield  {title} {\bibinfo {title} {Exploring materials band structure space with unsupervised machine learning},\ }\href@noop {} {\bibfield  {journal} {\bibinfo  {journal} {Computational Materials Science}\ }\textbf {\bibinfo {volume} {158}},\ \bibinfo {pages} {117} (\bibinfo {year} {2019})}\BibitemShut {NoStop}%
\bibitem [{\citenamefont {Scheurer}\ and\ \citenamefont {Slager}(2020)}]{feature_1}%
  \BibitemOpen
  \bibfield  {author} {\bibinfo {author} {\bibfnamefont {M.~S.}\ \bibnamefont {Scheurer}}\ and\ \bibinfo {author} {\bibfnamefont {R.-J.}\ \bibnamefont {Slager}},\ }\bibfield  {title} {\bibinfo {title} {Unsupervised machine learning and band topology},\ }\href@noop {} {\bibfield  {journal} {\bibinfo  {journal} {Physical review letters}\ }\textbf {\bibinfo {volume} {124}},\ \bibinfo {pages} {226401} (\bibinfo {year} {2020})}\BibitemShut {NoStop}%
\bibitem [{\citenamefont {Zimmermann}\ and\ \citenamefont {Jain}(2020)}]{Crystal}%
  \BibitemOpen
  \bibfield  {author} {\bibinfo {author} {\bibfnamefont {N.~E.}\ \bibnamefont {Zimmermann}}\ and\ \bibinfo {author} {\bibfnamefont {A.}~\bibnamefont {Jain}},\ }\bibfield  {title} {\bibinfo {title} {Local structure order parameters and site fingerprints for quantification of coordination environment and crystal structure similarity},\ }\href@noop {} {\bibfield  {journal} {\bibinfo  {journal} {RSC advances}\ }\textbf {\bibinfo {volume} {10}},\ \bibinfo {pages} {6063} (\bibinfo {year} {2020})}\BibitemShut {NoStop}%
\bibitem [{\citenamefont {Zhou}\ \emph {et~al.}(2019)\citenamefont {Zhou}, \citenamefont {Shen}, \citenamefont {Costa}, \citenamefont {Persson}, \citenamefont {Ong}, \citenamefont {Huck}, \citenamefont {Lu}, \citenamefont {Ma}, \citenamefont {Chen}, \citenamefont {Tang} \emph {et~al.}}]{2Dmatpedia}%
  \BibitemOpen
  \bibfield  {author} {\bibinfo {author} {\bibfnamefont {J.}~\bibnamefont {Zhou}}, \bibinfo {author} {\bibfnamefont {L.}~\bibnamefont {Shen}}, \bibinfo {author} {\bibfnamefont {M.~D.}\ \bibnamefont {Costa}}, \bibinfo {author} {\bibfnamefont {K.~A.}\ \bibnamefont {Persson}}, \bibinfo {author} {\bibfnamefont {S.~P.}\ \bibnamefont {Ong}}, \bibinfo {author} {\bibfnamefont {P.}~\bibnamefont {Huck}}, \bibinfo {author} {\bibfnamefont {Y.}~\bibnamefont {Lu}}, \bibinfo {author} {\bibfnamefont {X.}~\bibnamefont {Ma}}, \bibinfo {author} {\bibfnamefont {Y.}~\bibnamefont {Chen}}, \bibinfo {author} {\bibfnamefont {H.}~\bibnamefont {Tang}}, \emph {et~al.},\ }\bibfield  {title} {\bibinfo {title} {{2DMatPedia, an open computational database of two-dimensional materials from top-down and bottom-up approaches}},\ }\href@noop {} {\bibfield  {journal} {\bibinfo  {journal} {Scientific data}\ }\textbf {\bibinfo {volume} {6}},\ \bibinfo {pages} {86} (\bibinfo {year} {2019})}\BibitemShut {NoStop}%
\bibitem [{\citenamefont {Checkelsky}\ \emph {et~al.}(2024)\citenamefont {Checkelsky}, \citenamefont {Bernevig}, \citenamefont {Coleman}, \citenamefont {Si},\ and\ \citenamefont {Paschen}}]{checkelsky2024flat}%
  \BibitemOpen
  \bibfield  {author} {\bibinfo {author} {\bibfnamefont {J.~G.}\ \bibnamefont {Checkelsky}}, \bibinfo {author} {\bibfnamefont {B.~A.}\ \bibnamefont {Bernevig}}, \bibinfo {author} {\bibfnamefont {P.}~\bibnamefont {Coleman}}, \bibinfo {author} {\bibfnamefont {Q.}~\bibnamefont {Si}},\ and\ \bibinfo {author} {\bibfnamefont {S.}~\bibnamefont {Paschen}},\ }\bibfield  {title} {\bibinfo {title} {Flat bands, strange metals and the kondo effect},\ }\href@noop {} {\bibfield  {journal} {\bibinfo  {journal} {Nature Reviews Materials}\ } (\bibinfo {year} {2024})}\BibitemShut {NoStop}%
\bibitem [{\citenamefont {Leykam}\ \emph {et~al.}(2018)\citenamefont {Leykam}, \citenamefont {Andreanov},\ and\ \citenamefont {Flach}}]{leykam2018artificial}%
  \BibitemOpen
  \bibfield  {author} {\bibinfo {author} {\bibfnamefont {D.}~\bibnamefont {Leykam}}, \bibinfo {author} {\bibfnamefont {A.}~\bibnamefont {Andreanov}},\ and\ \bibinfo {author} {\bibfnamefont {S.}~\bibnamefont {Flach}},\ }\bibfield  {title} {\bibinfo {title} {Artificial flat band systems: from lattice models to experiments},\ }\href@noop {} {\bibfield  {journal} {\bibinfo  {journal} {Advances in Physics: X}\ }\textbf {\bibinfo {volume} {3}},\ \bibinfo {pages} {1473052} (\bibinfo {year} {2018})}\BibitemShut {NoStop}%
\bibitem [{\citenamefont {T{\"o}rm{\"a}}\ \emph {et~al.}(2022)\citenamefont {T{\"o}rm{\"a}}, \citenamefont {Peotta},\ and\ \citenamefont {Bernevig}}]{torma2022superconductivity}%
  \BibitemOpen
  \bibfield  {author} {\bibinfo {author} {\bibfnamefont {P.}~\bibnamefont {T{\"o}rm{\"a}}}, \bibinfo {author} {\bibfnamefont {S.}~\bibnamefont {Peotta}},\ and\ \bibinfo {author} {\bibfnamefont {B.~A.}\ \bibnamefont {Bernevig}},\ }\bibfield  {title} {\bibinfo {title} {Superconductivity, superfluidity and quantum geometry in twisted multilayer systems},\ }\href@noop {} {\bibfield  {journal} {\bibinfo  {journal} {Nature Reviews Physics}\ }\textbf {\bibinfo {volume} {4}},\ \bibinfo {pages} {528} (\bibinfo {year} {2022})}\BibitemShut {NoStop}%
\bibitem [{\citenamefont {Rhim}\ and\ \citenamefont {Yang}(2021)}]{rhim2021singular}%
  \BibitemOpen
  \bibfield  {author} {\bibinfo {author} {\bibfnamefont {J.-W.}\ \bibnamefont {Rhim}}\ and\ \bibinfo {author} {\bibfnamefont {B.-J.}\ \bibnamefont {Yang}},\ }\bibfield  {title} {\bibinfo {title} {Singular flat bands},\ }\href@noop {} {\bibfield  {journal} {\bibinfo  {journal} {Advances in Physics: X}\ }\textbf {\bibinfo {volume} {6}},\ \bibinfo {pages} {1901606} (\bibinfo {year} {2021})}\BibitemShut {NoStop}%
\bibitem [{\citenamefont {Huang}\ \emph {et~al.}(2022)\citenamefont {Huang}, \citenamefont {Tu}, \citenamefont {Shen}, \citenamefont {Zheng}, \citenamefont {Wang}, \citenamefont {Wang}, \citenamefont {Khaliji}, \citenamefont {Park}, \citenamefont {Liu}, \citenamefont {Yang} \emph {et~al.}}]{plasmons}%
  \BibitemOpen
  \bibfield  {author} {\bibinfo {author} {\bibfnamefont {T.}~\bibnamefont {Huang}}, \bibinfo {author} {\bibfnamefont {X.}~\bibnamefont {Tu}}, \bibinfo {author} {\bibfnamefont {C.}~\bibnamefont {Shen}}, \bibinfo {author} {\bibfnamefont {B.}~\bibnamefont {Zheng}}, \bibinfo {author} {\bibfnamefont {J.}~\bibnamefont {Wang}}, \bibinfo {author} {\bibfnamefont {H.}~\bibnamefont {Wang}}, \bibinfo {author} {\bibfnamefont {K.}~\bibnamefont {Khaliji}}, \bibinfo {author} {\bibfnamefont {S.~H.}\ \bibnamefont {Park}}, \bibinfo {author} {\bibfnamefont {Z.}~\bibnamefont {Liu}}, \bibinfo {author} {\bibfnamefont {T.}~\bibnamefont {Yang}}, \emph {et~al.},\ }\bibfield  {title} {\bibinfo {title} {Observation of chiral and slow plasmons in twisted bilayer graphene},\ }\href@noop {} {\bibfield  {journal} {\bibinfo  {journal} {Nature}\ }\textbf {\bibinfo {volume} {605}},\ \bibinfo {pages} {63} (\bibinfo {year} {2022})}\BibitemShut {NoStop}%
\bibitem [{\citenamefont {Cao}\ \emph {et~al.}(2018)\citenamefont {Cao}, \citenamefont {Fatemi}, \citenamefont {Fang}, \citenamefont {Watanabe}, \citenamefont {Taniguchi}, \citenamefont {Kaxiras},\ and\ \citenamefont {Jarillo-Herrero}}]{twist}%
  \BibitemOpen
  \bibfield  {author} {\bibinfo {author} {\bibfnamefont {Y.}~\bibnamefont {Cao}}, \bibinfo {author} {\bibfnamefont {V.}~\bibnamefont {Fatemi}}, \bibinfo {author} {\bibfnamefont {S.}~\bibnamefont {Fang}}, \bibinfo {author} {\bibfnamefont {K.}~\bibnamefont {Watanabe}}, \bibinfo {author} {\bibfnamefont {T.}~\bibnamefont {Taniguchi}}, \bibinfo {author} {\bibfnamefont {E.}~\bibnamefont {Kaxiras}},\ and\ \bibinfo {author} {\bibfnamefont {P.}~\bibnamefont {Jarillo-Herrero}},\ }\bibfield  {title} {\bibinfo {title} {Magic-angle graphene superlattices: a new platform for unconventional superconductivity},\ }\href@noop {} {\bibfield  {journal} {\bibinfo  {journal} {arXiv: 1803.02342}\ } (\bibinfo {year} {2018})}\BibitemShut {NoStop}%
\bibitem [{\citenamefont {Choi}\ \emph {et~al.}(2021)\citenamefont {Choi}, \citenamefont {Kim}, \citenamefont {Peng}, \citenamefont {Thomson}, \citenamefont {Lewandowski}, \citenamefont {Polski}, \citenamefont {Zhang}, \citenamefont {Arora}, \citenamefont {Watanabe}, \citenamefont {Taniguchi} \emph {et~al.}}]{Chern}%
  \BibitemOpen
  \bibfield  {author} {\bibinfo {author} {\bibfnamefont {Y.}~\bibnamefont {Choi}}, \bibinfo {author} {\bibfnamefont {H.}~\bibnamefont {Kim}}, \bibinfo {author} {\bibfnamefont {Y.}~\bibnamefont {Peng}}, \bibinfo {author} {\bibfnamefont {A.}~\bibnamefont {Thomson}}, \bibinfo {author} {\bibfnamefont {C.}~\bibnamefont {Lewandowski}}, \bibinfo {author} {\bibfnamefont {R.}~\bibnamefont {Polski}}, \bibinfo {author} {\bibfnamefont {Y.}~\bibnamefont {Zhang}}, \bibinfo {author} {\bibfnamefont {H.~S.}\ \bibnamefont {Arora}}, \bibinfo {author} {\bibfnamefont {K.}~\bibnamefont {Watanabe}}, \bibinfo {author} {\bibfnamefont {T.}~\bibnamefont {Taniguchi}}, \emph {et~al.},\ }\bibfield  {title} {\bibinfo {title} {Correlation-driven topological phases in magic-angle twisted bilayer graphene},\ }\href@noop {} {\bibfield  {journal} {\bibinfo  {journal} {Nature}\ }\textbf {\bibinfo {volume} {589}},\ \bibinfo {pages} {536} (\bibinfo {year} {2021})}\BibitemShut {NoStop}%
\bibitem [{\citenamefont {Duan}\ \emph {et~al.}(2024)\citenamefont {Duan}, \citenamefont {Ma}, \citenamefont {Zhang}, \citenamefont {Jiang}, \citenamefont {Zhang}, \citenamefont {Cui}, \citenamefont {Yu},\ and\ \citenamefont {Yao}}]{2dnewcatelogue}%
  \BibitemOpen
  \bibfield  {author} {\bibinfo {author} {\bibfnamefont {J.}~\bibnamefont {Duan}}, \bibinfo {author} {\bibfnamefont {D.-S.}\ \bibnamefont {Ma}}, \bibinfo {author} {\bibfnamefont {R.-W.}\ \bibnamefont {Zhang}}, \bibinfo {author} {\bibfnamefont {W.}~\bibnamefont {Jiang}}, \bibinfo {author} {\bibfnamefont {Z.}~\bibnamefont {Zhang}}, \bibinfo {author} {\bibfnamefont {C.}~\bibnamefont {Cui}}, \bibinfo {author} {\bibfnamefont {Z.-M.}\ \bibnamefont {Yu}},\ and\ \bibinfo {author} {\bibfnamefont {Y.}~\bibnamefont {Yao}},\ }\bibfield  {title} {\bibinfo {title} {Cataloging high-quality two-dimensional van der {Waals} materials with flat bands},\ }\href@noop {} {\bibfield  {journal} {\bibinfo  {journal} {Advanced Functional Materials}\ }\textbf {\bibinfo {volume} {34}},\ \bibinfo {pages} {2313067} (\bibinfo {year} {2024})}\BibitemShut {NoStop}%
\bibitem [{\citenamefont {Neves}\ \emph {et~al.}(2024)\citenamefont {Neves}, \citenamefont {Wakefield}, \citenamefont {Fang}, \citenamefont {Nguyen}, \citenamefont {Ye},\ and\ \citenamefont {Checkelsky}}]{crystalnet}%
  \BibitemOpen
  \bibfield  {author} {\bibinfo {author} {\bibfnamefont {P.~M.}\ \bibnamefont {Neves}}, \bibinfo {author} {\bibfnamefont {J.~P.}\ \bibnamefont {Wakefield}}, \bibinfo {author} {\bibfnamefont {S.}~\bibnamefont {Fang}}, \bibinfo {author} {\bibfnamefont {H.}~\bibnamefont {Nguyen}}, \bibinfo {author} {\bibfnamefont {L.}~\bibnamefont {Ye}},\ and\ \bibinfo {author} {\bibfnamefont {J.~G.}\ \bibnamefont {Checkelsky}},\ }\bibfield  {title} {\bibinfo {title} {Crystal net catalog of model flat band materials},\ }\href@noop {} {\bibfield  {journal} {\bibinfo  {journal} {npj Computational Materials}\ }\textbf {\bibinfo {volume} {10}},\ \bibinfo {pages} {39} (\bibinfo {year} {2024})}\BibitemShut {NoStop}%
\bibitem [{\citenamefont {Regnault}\ \emph {et~al.}(2022)\citenamefont {Regnault}, \citenamefont {Xu}, \citenamefont {Li}, \citenamefont {Ma}, \citenamefont {Jovanovic}, \citenamefont {Yazdani}, \citenamefont {Parkin}, \citenamefont {Felser}, \citenamefont {Schoop}, \citenamefont {Ong} \emph {et~al.}}]{flat_Cat}%
  \BibitemOpen
  \bibfield  {author} {\bibinfo {author} {\bibfnamefont {N.}~\bibnamefont {Regnault}}, \bibinfo {author} {\bibfnamefont {Y.}~\bibnamefont {Xu}}, \bibinfo {author} {\bibfnamefont {M.-R.}\ \bibnamefont {Li}}, \bibinfo {author} {\bibfnamefont {D.-S.}\ \bibnamefont {Ma}}, \bibinfo {author} {\bibfnamefont {M.}~\bibnamefont {Jovanovic}}, \bibinfo {author} {\bibfnamefont {A.}~\bibnamefont {Yazdani}}, \bibinfo {author} {\bibfnamefont {S.~S.}\ \bibnamefont {Parkin}}, \bibinfo {author} {\bibfnamefont {C.}~\bibnamefont {Felser}}, \bibinfo {author} {\bibfnamefont {L.~M.}\ \bibnamefont {Schoop}}, \bibinfo {author} {\bibfnamefont {N.~P.}\ \bibnamefont {Ong}}, \emph {et~al.},\ }\bibfield  {title} {\bibinfo {title} {Catalogue of flat-band stoichiometric materials},\ }\href@noop {} {\bibfield  {journal} {\bibinfo  {journal} {Nature}\ }\textbf {\bibinfo {volume} {603}},\ \bibinfo {pages} {824} (\bibinfo {year} {2022})}\BibitemShut {NoStop}%
\bibitem [{\citenamefont {Zhang}\ \emph {et~al.}(2023)\citenamefont {Zhang}, \citenamefont {Zhao}, \citenamefont {Song},\ and\ \citenamefont {Shen}}]{zhang2023physically}%
  \BibitemOpen
  \bibfield  {author} {\bibinfo {author} {\bibfnamefont {X.}~\bibnamefont {Zhang}}, \bibinfo {author} {\bibfnamefont {Y.-M.}\ \bibnamefont {Zhao}}, \bibinfo {author} {\bibfnamefont {Z.}~\bibnamefont {Song}},\ and\ \bibinfo {author} {\bibfnamefont {L.}~\bibnamefont {Shen}},\ }\bibfield  {title} {\bibinfo {title} {Physically explainable statistical learning of flat bands in stoichiometric materials from the periodic table},\ }\href@noop {} {\bibfield  {journal} {\bibinfo  {journal} {Physical Review Materials}\ }\textbf {\bibinfo {volume} {7}},\ \bibinfo {pages} {064804} (\bibinfo {year} {2023})}\BibitemShut {NoStop}%
\bibitem [{\citenamefont {Campello}\ \emph {et~al.}(2013)\citenamefont {Campello}, \citenamefont {Moulavi},\ and\ \citenamefont {Sander}}]{HDB}%
  \BibitemOpen
  \bibfield  {author} {\bibinfo {author} {\bibfnamefont {R.~J. G.~B.}\ \bibnamefont {Campello}}, \bibinfo {author} {\bibfnamefont {D.}~\bibnamefont {Moulavi}},\ and\ \bibinfo {author} {\bibfnamefont {J.}~\bibnamefont {Sander}},\ }\bibfield  {title} {\bibinfo {title} {Density-based clustering based on hierarchical density estimates},\ }in\ \href@noop {} {\emph {\bibinfo {booktitle} {Advances in Knowledge Discovery and Data Mining}}}\ (\bibinfo {year} {2013})\ pp.\ \bibinfo {pages} {160--172}\BibitemShut {NoStop}%
\bibitem [{\citenamefont {Van~der Maaten}\ and\ \citenamefont {Hinton}(2008)}]{t-sne}%
  \BibitemOpen
  \bibfield  {author} {\bibinfo {author} {\bibfnamefont {L.}~\bibnamefont {Van~der Maaten}}\ and\ \bibinfo {author} {\bibfnamefont {G.}~\bibnamefont {Hinton}},\ }\bibfield  {title} {\bibinfo {title} {Visualizing data using {t-SNE}.},\ }\href@noop {} {\bibfield  {journal} {\bibinfo  {journal} {Journal of machine learning research}\ }\textbf {\bibinfo {volume} {9}} (\bibinfo {year} {2008})}\BibitemShut {NoStop}%
\bibitem [{\citenamefont {Ester}\ \emph {et~al.}(1996)\citenamefont {Ester}, \citenamefont {Kriegel}, \citenamefont {Sander},\ and\ \citenamefont {Xu}}]{DBSCAN}%
  \BibitemOpen
  \bibfield  {author} {\bibinfo {author} {\bibfnamefont {M.}~\bibnamefont {Ester}}, \bibinfo {author} {\bibfnamefont {H.-P.}\ \bibnamefont {Kriegel}}, \bibinfo {author} {\bibfnamefont {J.}~\bibnamefont {Sander}},\ and\ \bibinfo {author} {\bibfnamefont {X.}~\bibnamefont {Xu}},\ }\bibfield  {title} {\bibinfo {title} {A density-based algorithm for discovering clusters in large spatial databases with noise},\ }in\ \href@noop {} {\emph {\bibinfo {booktitle} {Proceedings of the Second International Conference on Knowledge Discovery and Data Mining}}}\ (\bibinfo {year} {1996})\ p.\ \bibinfo {pages} {226–231}\BibitemShut {NoStop}%
\bibitem [{\citenamefont {Lanzillo}\ \emph {et~al.}(2015)\citenamefont {Lanzillo}, \citenamefont {Roy},\ and\ \citenamefont {Nayak}}]{Hg}%
  \BibitemOpen
  \bibfield  {author} {\bibinfo {author} {\bibfnamefont {N.~A.}\ \bibnamefont {Lanzillo}}, \bibinfo {author} {\bibfnamefont {S.}~\bibnamefont {Roy}},\ and\ \bibinfo {author} {\bibfnamefont {S.~K.}\ \bibnamefont {Nayak}},\ }\bibfield  {title} {\bibinfo {title} {Quantum confinement and quasiparticle corrections in {$\alpha$-HgS} from first principles},\ }\href@noop {} {\bibfield  {journal} {\bibinfo  {journal} {Surface Science}\ }\textbf {\bibinfo {volume} {636}},\ \bibinfo {pages} {54} (\bibinfo {year} {2015})}\BibitemShut {NoStop}%
\bibitem [{\citenamefont {Li}\ \emph {et~al.}(2015)\citenamefont {Li}, \citenamefont {He}, \citenamefont {Meng}, \citenamefont {Xiao}, \citenamefont {Tang}, \citenamefont {Wei}, \citenamefont {Kim}, \citenamefont {Kioussis}, \citenamefont {Malcolm~Stocks},\ and\ \citenamefont {Zhong}}]{Hg_2}%
  \BibitemOpen
  \bibfield  {author} {\bibinfo {author} {\bibfnamefont {J.}~\bibnamefont {Li}}, \bibinfo {author} {\bibfnamefont {C.}~\bibnamefont {He}}, \bibinfo {author} {\bibfnamefont {L.}~\bibnamefont {Meng}}, \bibinfo {author} {\bibfnamefont {H.}~\bibnamefont {Xiao}}, \bibinfo {author} {\bibfnamefont {C.}~\bibnamefont {Tang}}, \bibinfo {author} {\bibfnamefont {X.}~\bibnamefont {Wei}}, \bibinfo {author} {\bibfnamefont {J.}~\bibnamefont {Kim}}, \bibinfo {author} {\bibfnamefont {N.}~\bibnamefont {Kioussis}}, \bibinfo {author} {\bibfnamefont {G.}~\bibnamefont {Malcolm~Stocks}},\ and\ \bibinfo {author} {\bibfnamefont {J.}~\bibnamefont {Zhong}},\ }\bibfield  {title} {\bibinfo {title} {Two-dimensional topological insulators with tunable band gaps: Single-layer hgte and hgse},\ }\href@noop {} {\bibfield  {journal} {\bibinfo  {journal} {Scientific reports}\ }\textbf {\bibinfo {volume} {5}},\ \bibinfo {pages} {14115} (\bibinfo {year} {2015})}\BibitemShut {NoStop}%
\bibitem [{\citenamefont {Imran}\ \emph {et~al.}(2018)\citenamefont {Imran}, \citenamefont {Co{\c{s}}kun}, \citenamefont {Isikgor}, \citenamefont {Bichen}, \citenamefont {Khan},\ and\ \citenamefont {Ouyang}}]{Zn}%
  \BibitemOpen
  \bibfield  {author} {\bibinfo {author} {\bibfnamefont {M.}~\bibnamefont {Imran}}, \bibinfo {author} {\bibfnamefont {H.}~\bibnamefont {Co{\c{s}}kun}}, \bibinfo {author} {\bibfnamefont {F.~H.}\ \bibnamefont {Isikgor}}, \bibinfo {author} {\bibfnamefont {L.}~\bibnamefont {Bichen}}, \bibinfo {author} {\bibfnamefont {N.~A.}\ \bibnamefont {Khan}},\ and\ \bibinfo {author} {\bibfnamefont {J.}~\bibnamefont {Ouyang}},\ }\bibfield  {title} {\bibinfo {title} {Highly efficient and stable inverted perovskite solar cells with two-dimensional znse deposited using a thermal evaporator for electron collection},\ }\href@noop {} {\bibfield  {journal} {\bibinfo  {journal} {Journal of materials chemistry A}\ }\textbf {\bibinfo {volume} {6}},\ \bibinfo {pages} {22713} (\bibinfo {year} {2018})}\BibitemShut {NoStop}%
\bibitem [{\citenamefont {Xiong}\ and\ \citenamefont {Zhou}(2019)}]{Zn_2}%
  \BibitemOpen
  \bibfield  {author} {\bibinfo {author} {\bibfnamefont {A.}~\bibnamefont {Xiong}}\ and\ \bibinfo {author} {\bibfnamefont {X.}~\bibnamefont {Zhou}},\ }\bibfield  {title} {\bibinfo {title} {Tunable electronic and optical properties of novel{ ZnSe/AlP van der Waals} heterostructure},\ }\href@noop {} {\bibfield  {journal} {\bibinfo  {journal} {Materials Research Express}\ }\textbf {\bibinfo {volume} {6}},\ \bibinfo {pages} {075907} (\bibinfo {year} {2019})}\BibitemShut {NoStop}%
\bibitem [{\citenamefont {Zhang}\ \emph {et~al.}(2020)\citenamefont {Zhang}, \citenamefont {Sun}, \citenamefont {Ye}, \citenamefont {Song},\ and\ \citenamefont {Qu}}]{Cd}%
  \BibitemOpen
  \bibfield  {author} {\bibinfo {author} {\bibfnamefont {J.}~\bibnamefont {Zhang}}, \bibinfo {author} {\bibfnamefont {Y.}~\bibnamefont {Sun}}, \bibinfo {author} {\bibfnamefont {S.}~\bibnamefont {Ye}}, \bibinfo {author} {\bibfnamefont {J.}~\bibnamefont {Song}},\ and\ \bibinfo {author} {\bibfnamefont {J.}~\bibnamefont {Qu}},\ }\bibfield  {title} {\bibinfo {title} {Heterostructures in two-dimensional {CdSe} nanoplatelets: synthesis, optical properties, and applications},\ }\href@noop {} {\bibfield  {journal} {\bibinfo  {journal} {Chemistry of Materials}\ }\textbf {\bibinfo {volume} {32}},\ \bibinfo {pages} {9490} (\bibinfo {year} {2020})}\BibitemShut {NoStop}%
\bibitem [{\citenamefont {Zhou}\ \emph {et~al.}(2022)\citenamefont {Zhou}, \citenamefont {Duo}, \citenamefont {Wang}, \citenamefont {Chu}, \citenamefont {Zhang},\ and\ \citenamefont {Yan}}]{Bi1}%
  \BibitemOpen
  \bibfield  {author} {\bibinfo {author} {\bibfnamefont {J.}~\bibnamefont {Zhou}}, \bibinfo {author} {\bibfnamefont {F.}~\bibnamefont {Duo}}, \bibinfo {author} {\bibfnamefont {C.}~\bibnamefont {Wang}}, \bibinfo {author} {\bibfnamefont {L.}~\bibnamefont {Chu}}, \bibinfo {author} {\bibfnamefont {M.}~\bibnamefont {Zhang}},\ and\ \bibinfo {author} {\bibfnamefont {D.}~\bibnamefont {Yan}},\ }\bibfield  {title} {\bibinfo {title} {Robust photocatalytic activity of two-dimensional {h-BN/Bi2O3} heterostructure quantum sheets},\ }\href@noop {} {\bibfield  {journal} {\bibinfo  {journal} {RSC advances}\ }\textbf {\bibinfo {volume} {12}},\ \bibinfo {pages} {13535} (\bibinfo {year} {2022})}\BibitemShut {NoStop}%
\bibitem [{\citenamefont {Mei}\ \emph {et~al.}(2019)\citenamefont {Mei}, \citenamefont {Liao}, \citenamefont {Ayoko},\ and\ \citenamefont {Sun}}]{Bi2}%
  \BibitemOpen
  \bibfield  {author} {\bibinfo {author} {\bibfnamefont {J.}~\bibnamefont {Mei}}, \bibinfo {author} {\bibfnamefont {T.}~\bibnamefont {Liao}}, \bibinfo {author} {\bibfnamefont {G.~A.}\ \bibnamefont {Ayoko}},\ and\ \bibinfo {author} {\bibfnamefont {Z.}~\bibnamefont {Sun}},\ }\bibfield  {title} {\bibinfo {title} {Two-dimensional bismuth oxide heterostructured nanosheets for lithium-and sodium-ion storages},\ }\href@noop {} {\bibfield  {journal} {\bibinfo  {journal} {ACS applied materials \& interfaces}\ }\textbf {\bibinfo {volume} {11}},\ \bibinfo {pages} {28205} (\bibinfo {year} {2019})}\BibitemShut {NoStop}%
\bibitem [{\citenamefont {Yu}\ \emph {et~al.}(2023)\citenamefont {Yu}, \citenamefont {Li}, \citenamefont {Wu}, \citenamefont {Lu},\ and\ \citenamefont {Zhang}}]{InAs}%
  \BibitemOpen
  \bibfield  {author} {\bibinfo {author} {\bibfnamefont {W.}~\bibnamefont {Yu}}, \bibinfo {author} {\bibfnamefont {J.}~\bibnamefont {Li}}, \bibinfo {author} {\bibfnamefont {Y.}~\bibnamefont {Wu}}, \bibinfo {author} {\bibfnamefont {J.}~\bibnamefont {Lu}},\ and\ \bibinfo {author} {\bibfnamefont {Y.}~\bibnamefont {Zhang}},\ }\bibfield  {title} {\bibinfo {title} {Systematic investigation of the mechanical, electronic, and interfacial properties of high mobility monolayer {InAs} from first-principles calculations},\ }\href@noop {} {\bibfield  {journal} {\bibinfo  {journal} {Physical Chemistry Chemical Physics}\ }\textbf {\bibinfo {volume} {25}},\ \bibinfo {pages} {10769} (\bibinfo {year} {2023})}\BibitemShut {NoStop}%
\bibitem [{\citenamefont {Wu}\ \emph {et~al.}(2020)\citenamefont {Wu}, \citenamefont {Chen}, \citenamefont {Ma}, \citenamefont {Wan}, \citenamefont {Hu},\ and\ \citenamefont {Yang}}]{AlBi}%
  \BibitemOpen
  \bibfield  {author} {\bibinfo {author} {\bibfnamefont {K.}~\bibnamefont {Wu}}, \bibinfo {author} {\bibfnamefont {J.}~\bibnamefont {Chen}}, \bibinfo {author} {\bibfnamefont {H.}~\bibnamefont {Ma}}, \bibinfo {author} {\bibfnamefont {L.}~\bibnamefont {Wan}}, \bibinfo {author} {\bibfnamefont {W.}~\bibnamefont {Hu}},\ and\ \bibinfo {author} {\bibfnamefont {J.}~\bibnamefont {Yang}},\ }\bibfield  {title} {\bibinfo {title} {Two-dimensional giant tunable rashba semiconductors with two-atom-thick buckled honeycomb structure},\ }\href@noop {} {\bibfield  {journal} {\bibinfo  {journal} {Nano Letters}\ }\textbf {\bibinfo {volume} {21}},\ \bibinfo {pages} {740} (\bibinfo {year} {2020})}\BibitemShut {NoStop}%
\bibitem [{\citenamefont {Wunderlich}\ \emph {et~al.}(2023)\citenamefont {Wunderlich}, \citenamefont {Ferrari},\ and\ \citenamefont {Valent{\'\i}}}]{Topol_2}%
  \BibitemOpen
  \bibfield  {author} {\bibinfo {author} {\bibfnamefont {P.}~\bibnamefont {Wunderlich}}, \bibinfo {author} {\bibfnamefont {F.}~\bibnamefont {Ferrari}},\ and\ \bibinfo {author} {\bibfnamefont {R.}~\bibnamefont {Valent{\'\i}}},\ }\bibfield  {title} {\bibinfo {title} {Detecting topological phases in the square--octagon lattice with statistical methods},\ }\href@noop {} {\bibfield  {journal} {\bibinfo  {journal} {The European Physical Journal Plus}\ }\textbf {\bibinfo {volume} {138}},\ \bibinfo {pages} {336} (\bibinfo {year} {2023})}\BibitemShut {NoStop}%
\bibitem [{\citenamefont {Bao}\ \emph {et~al.}(2014)\citenamefont {Bao}, \citenamefont {Tao}, \citenamefont {Liu}, \citenamefont {Zhang},\ and\ \citenamefont {Liu}}]{Topol}%
  \BibitemOpen
  \bibfield  {author} {\bibinfo {author} {\bibfnamefont {A.}~\bibnamefont {Bao}}, \bibinfo {author} {\bibfnamefont {H.-S.}\ \bibnamefont {Tao}}, \bibinfo {author} {\bibfnamefont {H.-D.}\ \bibnamefont {Liu}}, \bibinfo {author} {\bibfnamefont {X.}~\bibnamefont {Zhang}},\ and\ \bibinfo {author} {\bibfnamefont {W.-M.}\ \bibnamefont {Liu}},\ }\bibfield  {title} {\bibinfo {title} {Quantum magnetic phase transition in square-octagon lattice},\ }\href@noop {} {\bibfield  {journal} {\bibinfo  {journal} {Scientific reports}\ }\textbf {\bibinfo {volume} {4}},\ \bibinfo {pages} {6918} (\bibinfo {year} {2014})}\BibitemShut {NoStop}%
\bibitem [{\citenamefont {Nunes}\ and\ \citenamefont {Smith}(2020)}]{SqOct}%
  \BibitemOpen
  \bibfield  {author} {\bibinfo {author} {\bibfnamefont {L.~H.}\ \bibnamefont {Nunes}}\ and\ \bibinfo {author} {\bibfnamefont {C.~M.}\ \bibnamefont {Smith}},\ }\bibfield  {title} {\bibinfo {title} {Flat-band superconductivity for tight-binding electrons on a square-octagon lattice},\ }\href@noop {} {\bibfield  {journal} {\bibinfo  {journal} {Physical Review B}\ }\textbf {\bibinfo {volume} {101}},\ \bibinfo {pages} {224514} (\bibinfo {year} {2020})}\BibitemShut {NoStop}%
\bibitem [{\citenamefont {Peotta}\ \emph {et~al.}(2023)\citenamefont {Peotta}, \citenamefont {Huhtinen},\ and\ \citenamefont {T{\"o}rm{\"a}}}]{flat_band_overlap}%
  \BibitemOpen
  \bibfield  {author} {\bibinfo {author} {\bibfnamefont {S.}~\bibnamefont {Peotta}}, \bibinfo {author} {\bibfnamefont {K.-E.}\ \bibnamefont {Huhtinen}},\ and\ \bibinfo {author} {\bibfnamefont {P.}~\bibnamefont {T{\"o}rm{\"a}}},\ }\bibfield  {title} {\bibinfo {title} {Quantum geometry in superfluidity and superconductivity},\ }\href@noop {} {\bibfield  {journal} {\bibinfo  {journal} {arXiv: 2308.08248}\ } (\bibinfo {year} {2023})}\BibitemShut {NoStop}%
\bibitem [{\citenamefont {Ong}\ \emph {et~al.}(2013)\citenamefont {Ong}, \citenamefont {Richards}, \citenamefont {Jain}, \citenamefont {Hautier}, \citenamefont {Kocher}, \citenamefont {Cholia}, \citenamefont {Gunter}, \citenamefont {Chevrier}, \citenamefont {Persson},\ and\ \citenamefont {Ceder}}]{pymatgen}%
  \BibitemOpen
  \bibfield  {author} {\bibinfo {author} {\bibfnamefont {S.~P.}\ \bibnamefont {Ong}}, \bibinfo {author} {\bibfnamefont {W.~D.}\ \bibnamefont {Richards}}, \bibinfo {author} {\bibfnamefont {A.}~\bibnamefont {Jain}}, \bibinfo {author} {\bibfnamefont {G.}~\bibnamefont {Hautier}}, \bibinfo {author} {\bibfnamefont {M.}~\bibnamefont {Kocher}}, \bibinfo {author} {\bibfnamefont {S.}~\bibnamefont {Cholia}}, \bibinfo {author} {\bibfnamefont {D.}~\bibnamefont {Gunter}}, \bibinfo {author} {\bibfnamefont {V.~L.}\ \bibnamefont {Chevrier}}, \bibinfo {author} {\bibfnamefont {K.~A.}\ \bibnamefont {Persson}},\ and\ \bibinfo {author} {\bibfnamefont {G.}~\bibnamefont {Ceder}},\ }\bibfield  {title} {\bibinfo {title} {Python materials genomics (pymatgen): A robust, open-source python library for materials analysis},\ }\href@noop {} {\bibfield  {journal} {\bibinfo  {journal} {Computational Materials Science}\ }\textbf {\bibinfo {volume} {68}},\ \bibinfo {pages} {314} (\bibinfo {year} {2013})}\BibitemShut {NoStop}%
\bibitem [{\citenamefont {Isayev}\ \emph {et~al.}(2015)\citenamefont {Isayev}, \citenamefont {Fourches}, \citenamefont {Muratov}, \citenamefont {Oses}, \citenamefont {Rasch}, \citenamefont {Tropsha},\ and\ \citenamefont {Curtarolo}}]{E_2}%
  \BibitemOpen
  \bibfield  {author} {\bibinfo {author} {\bibfnamefont {O.}~\bibnamefont {Isayev}}, \bibinfo {author} {\bibfnamefont {D.}~\bibnamefont {Fourches}}, \bibinfo {author} {\bibfnamefont {E.~N.}\ \bibnamefont {Muratov}}, \bibinfo {author} {\bibfnamefont {C.}~\bibnamefont {Oses}}, \bibinfo {author} {\bibfnamefont {K.}~\bibnamefont {Rasch}}, \bibinfo {author} {\bibfnamefont {A.}~\bibnamefont {Tropsha}},\ and\ \bibinfo {author} {\bibfnamefont {S.}~\bibnamefont {Curtarolo}},\ }\bibfield  {title} {\bibinfo {title} {Materials cartography: representing and mining materials space using structural and electronic fingerprints},\ }\href@noop {} {\bibfield  {journal} {\bibinfo  {journal} {Chemistry of Materials}\ }\textbf {\bibinfo {volume} {27}},\ \bibinfo {pages} {735} (\bibinfo {year} {2015})}\BibitemShut {NoStop}%
\bibitem [{\citenamefont {Kn{\o}sgaard}\ and\ \citenamefont {Thygesen}(2022)}]{E_1}%
  \BibitemOpen
  \bibfield  {author} {\bibinfo {author} {\bibfnamefont {N.~R.}\ \bibnamefont {Kn{\o}sgaard}}\ and\ \bibinfo {author} {\bibfnamefont {K.~S.}\ \bibnamefont {Thygesen}},\ }\bibfield  {title} {\bibinfo {title} {Representing individual electronic states for machine learning {GW band structures of 2D} materials},\ }\href@noop {} {\bibfield  {journal} {\bibinfo  {journal} {Nature Communications}\ }\textbf {\bibinfo {volume} {13}},\ \bibinfo {pages} {468} (\bibinfo {year} {2022})}\BibitemShut {NoStop}%
\bibitem [{\citenamefont {He}\ \emph {et~al.}(2016)\citenamefont {He}, \citenamefont {Zhang}, \citenamefont {Ren},\ and\ \citenamefont {Sun}}]{Res_11}%
  \BibitemOpen
  \bibfield  {author} {\bibinfo {author} {\bibfnamefont {K.}~\bibnamefont {He}}, \bibinfo {author} {\bibfnamefont {X.}~\bibnamefont {Zhang}}, \bibinfo {author} {\bibfnamefont {S.}~\bibnamefont {Ren}},\ and\ \bibinfo {author} {\bibfnamefont {J.}~\bibnamefont {Sun}},\ }\bibfield  {title} {\bibinfo {title} {Deep residual learning for image recognition},\ }in\ \href@noop {} {\emph {\bibinfo {booktitle} {Proceedings of the IEEE conference on computer vision and pattern recognition}}}\ (\bibinfo {year} {2016})\ pp.\ \bibinfo {pages} {770--778}\BibitemShut {NoStop}%
\bibitem [{\citenamefont {Horizon2333}(2021)}]{Res_22}%
  \BibitemOpen
  \bibfield  {author} {\bibinfo {author} {\bibnamefont {Horizon2333}},\ }\href@noop {} {\bibinfo {title} {Imagenet autoencoder}},\ \bibinfo {howpublished} {\url{https://github.com/Horizon2333/imagenet-autoencoder}} (\bibinfo {year} {2021}),\ \bibinfo {note} {{GitHub repository}}\BibitemShut {NoStop}%
\bibitem [{\citenamefont {Yu}\ and\ \citenamefont {Wang}(2019)}]{Res_1}%
  \BibitemOpen
  \bibfield  {author} {\bibinfo {author} {\bibfnamefont {X.}~\bibnamefont {Yu}}\ and\ \bibinfo {author} {\bibfnamefont {S.-H.}\ \bibnamefont {Wang}},\ }\bibfield  {title} {\bibinfo {title} {Abnormality diagnosis in mammograms by transfer learning based on resnet18},\ }\href@noop {} {\bibfield  {journal} {\bibinfo  {journal} {Fundamenta Informaticae}\ }\textbf {\bibinfo {volume} {168}},\ \bibinfo {pages} {219} (\bibinfo {year} {2019})}\BibitemShut {NoStop}%
\bibitem [{\citenamefont {Odusami}\ \emph {et~al.}(2021)\citenamefont {Odusami}, \citenamefont {Maskeli{\=u}nas}, \citenamefont {Dama{\v{s}}evi{\v{c}}ius},\ and\ \citenamefont {Krilavi{\v{c}}ius}}]{Res_2}%
  \BibitemOpen
  \bibfield  {author} {\bibinfo {author} {\bibfnamefont {M.}~\bibnamefont {Odusami}}, \bibinfo {author} {\bibfnamefont {R.}~\bibnamefont {Maskeli{\=u}nas}}, \bibinfo {author} {\bibfnamefont {R.}~\bibnamefont {Dama{\v{s}}evi{\v{c}}ius}},\ and\ \bibinfo {author} {\bibfnamefont {T.}~\bibnamefont {Krilavi{\v{c}}ius}},\ }\bibfield  {title} {\bibinfo {title} {Analysis of features of alzheimer’s disease: Detection of early stage from functional brain changes in magnetic resonance images using a finetuned resnet18 network},\ }\href@noop {} {\bibfield  {journal} {\bibinfo  {journal} {Diagnostics}\ }\textbf {\bibinfo {volume} {11}},\ \bibinfo {pages} {1071} (\bibinfo {year} {2021})}\BibitemShut {NoStop}%
\bibitem [{\citenamefont {Creswell}\ \emph {et~al.}(2017)\citenamefont {Creswell}, \citenamefont {Arulkumaran},\ and\ \citenamefont {Bharath}}]{BCE}%
  \BibitemOpen
  \bibfield  {author} {\bibinfo {author} {\bibfnamefont {A.}~\bibnamefont {Creswell}}, \bibinfo {author} {\bibfnamefont {K.}~\bibnamefont {Arulkumaran}},\ and\ \bibinfo {author} {\bibfnamefont {A.~A.}\ \bibnamefont {Bharath}},\ }\bibfield  {title} {\bibinfo {title} {On denoising autoencoders trained to minimise binary cross-entropy},\ }\href@noop {} {\bibfield  {journal} {\bibinfo  {journal} {arXiv: 1708.08487}\ } (\bibinfo {year} {2017})}\BibitemShut {NoStop}%
\bibitem [{\citenamefont {Roweis}\ and\ \citenamefont {Saul}(2000)}]{LLE}%
  \BibitemOpen
  \bibfield  {author} {\bibinfo {author} {\bibfnamefont {S.~T.}\ \bibnamefont {Roweis}}\ and\ \bibinfo {author} {\bibfnamefont {L.~K.}\ \bibnamefont {Saul}},\ }\bibfield  {title} {\bibinfo {title} {Nonlinear dimensionality reduction by locally linear embedding},\ }\href@noop {} {\bibfield  {journal} {\bibinfo  {journal} {science}\ }\textbf {\bibinfo {volume} {290}},\ \bibinfo {pages} {2323} (\bibinfo {year} {2000})}\BibitemShut {NoStop}%
\bibitem [{\citenamefont {Donoho}\ and\ \citenamefont {Grimes}(2003)}]{Hessian}%
  \BibitemOpen
  \bibfield  {author} {\bibinfo {author} {\bibfnamefont {D.~L.}\ \bibnamefont {Donoho}}\ and\ \bibinfo {author} {\bibfnamefont {C.}~\bibnamefont {Grimes}},\ }\bibfield  {title} {\bibinfo {title} {Hessian eigenmaps: Locally linear embedding techniques for high-dimensional data},\ }\href@noop {} {\bibfield  {journal} {\bibinfo  {journal} {Proceedings of the National Academy of Sciences}\ }\textbf {\bibinfo {volume} {100}},\ \bibinfo {pages} {5591} (\bibinfo {year} {2003})}\BibitemShut {NoStop}%
\bibitem [{\citenamefont {McInnes}\ \emph {et~al.}(2018)\citenamefont {McInnes}, \citenamefont {Healy},\ and\ \citenamefont {Melville}}]{UMAP}%
  \BibitemOpen
  \bibfield  {author} {\bibinfo {author} {\bibfnamefont {L.}~\bibnamefont {McInnes}}, \bibinfo {author} {\bibfnamefont {J.}~\bibnamefont {Healy}},\ and\ \bibinfo {author} {\bibfnamefont {J.}~\bibnamefont {Melville}},\ }\bibfield  {title} {\bibinfo {title} {Umap: Uniform manifold approximation and projection for dimension reduction},\ }\href@noop {} {\bibfield  {journal} {\bibinfo  {journal} {arXiv: 1802.03426}\ } (\bibinfo {year} {2018})}\BibitemShut {NoStop}%
\bibitem [{\citenamefont {Aggarwal}\ \emph {et~al.}(2001)\citenamefont {Aggarwal}, \citenamefont {Hinneburg},\ and\ \citenamefont {Keim}}]{Metric}%
  \BibitemOpen
  \bibfield  {author} {\bibinfo {author} {\bibfnamefont {C.~C.}\ \bibnamefont {Aggarwal}}, \bibinfo {author} {\bibfnamefont {A.}~\bibnamefont {Hinneburg}},\ and\ \bibinfo {author} {\bibfnamefont {D.~A.}\ \bibnamefont {Keim}},\ }\bibfield  {title} {\bibinfo {title} {On the surprising behavior of distance metrics in high dimensional space},\ }in\ \href@noop {} {\emph {\bibinfo {booktitle} {Database Theory --- ICDT}}}\ (\bibinfo {year} {2001})\ pp.\ \bibinfo {pages} {420--434}\BibitemShut {NoStop}%
\bibitem [{\citenamefont {Moulavi}\ \emph {et~al.}(2014)\citenamefont {Moulavi}, \citenamefont {Jaskowiak}, \citenamefont {Campello}, \citenamefont {Zimek},\ and\ \citenamefont {Sander}}]{DBCV}%
  \BibitemOpen
  \bibfield  {author} {\bibinfo {author} {\bibfnamefont {D.}~\bibnamefont {Moulavi}}, \bibinfo {author} {\bibfnamefont {P.~A.}\ \bibnamefont {Jaskowiak}}, \bibinfo {author} {\bibfnamefont {R.~J.}\ \bibnamefont {Campello}}, \bibinfo {author} {\bibfnamefont {A.}~\bibnamefont {Zimek}},\ and\ \bibinfo {author} {\bibfnamefont {J.}~\bibnamefont {Sander}},\ }\bibfield  {title} {\bibinfo {title} {Density-based clustering validation},\ }in\ \href@noop {} {\emph {\bibinfo {booktitle} {Proceedings of the 2014 SIAM international conference on data mining}}}\ (\bibinfo {year} {2014})\ pp.\ \bibinfo {pages} {839--847}\BibitemShut {NoStop}%
\end{thebibliography}%

\end{document}